\documentstyle[epsf,aps]{revtex}
\newenvironment{figures}[1]%
{\begin{list}{}{\settowidth{\labelwidth}{#1}
  \setlength{\leftmargin}{\labelwidth}
  \addtolength{\leftmargin}{\labelsep}
  \setlength{\parsep}{1ex plus0.7ex minus0.7ex}
  \setlength{\itemsep}{0.8ex}
  }}{\end{list}}

\def\mc#1 {\multicolumn{1}{|c|}{#1}}

\newcommand{\beq}{\begin{equation}}
\newcommand{\eeq}{\end{equation}}
\newcommand{\beqa}{\begin{eqnarray}}
\newcommand{\eeqa}{\end{eqnarray}}
\newcommand{\beqaT}{\begin{eqnarray}{ccc}}
\newcommand{\eeqaT}{\end{eqnarray}}

\begin{document}

\title {
{\small
\hspace{11cm} NIKHEF 97--015\\
\hspace{11cm} hep-ph/9704203 }\\ \vspace{1cm}
{\bf Charm production in deep inelastic and diffractive scattering}}

\author{
L.P.A. Haakman
$^{a}$,
A.B. Kaidalov
$^{b}$,
J.H. Koch
$^{a}$}
\address{
$^{a}$
National Institute for Nuclear Physics
and High Energy Physics (NIKHEF), \\
P.O. Box 41882, NL-1009 DB Amsterdam, The Netherlands\\
$^{b}$
Institute of Theoretical and Experimental Physics, \\
B. Cheremushinskaya 25, 117 259 Moscow, Russia}

\date{March 1997}

\maketitle

\begin{abstract}
We consider the production of charm by real and virtual photons.
Special attention is paid to diffractive charm production,
which provides information on the gluonic content of the Pomeron. 
Our calculations are based on the gluon distributions of the CKMT-model,
which is shown to lead to agreement with the data on open
charm production in deep inelastic scattering. 
We compare predictions for diffractive charm production 
of different models for the distribution of gluons in the Pomeron. 
Experiments at HERA should be able to discriminate between them.  
Predictions for beauty production in diffractive and 
non-diffractive interactions of photons are also given. 
\end{abstract}


\section{Introduction}
The production of heavy quarks in deep inelastic electron 
scattering from a proton has received increasing attention 
recently, since this reaction is seen as a tool to probe the 
gluon distribution in the nucleon.
Due to the large mass of the quark, the reaction is believed to
be driven by a perturbative mechanism, photon-gluon fusion,
and is therefore sensitive to the nucleonic gluon density. 
We consider here in a consistent way the production of charm 
in deep inelastic scattering (DIS) as well as in hard diffractive 
scattering. 
The latter process, while driven by the same basic reaction 
mechanism, probes a different gluon density.
As the diffractive processes are mostly seen as mediated by 
the Pomeron exchange, the relevant quantity in 
diffractive production is then the gluon distribution in the 
Pomeron; a review can be found in Ref.\cite{Kaidalov}. 

The distribution of gluons in the proton is known now comparatively 
well from deep inelatic scattering for $x\geq 10^{-2}$, but for 
smaller values of $x$ information on $g_p(x,Q^2)$ is very limited. 
A comparison of theoretical predictions on charm production 
in the HERA energy range with experimental data 
allows one to test small-$x$ behaviour of gluonic distributions.

Diffractive production of hadrons in deep inelastic
scattering has been observed in experiments at HERA \cite{ZEUSD,H1D}. 
The gluonic content of the Pomeron is poorly known. 
Studies of the $Q^2$-dependence of the Pomeron structure function \cite{CapellaFP1,CapellaFP2,Gehrman,Golec,H1P} 
lead to the conclusion that the distribution of gluons in the
Pomeron is hard and that they carry the main part of the Pomeron 
momentum. 
However there are big differences between gluon distribution 
functions $g_P(x,Q^2)$ in different models. 

It is the purpose of this paper to present a consistent approach to 
open charm photo- and electroproduction as well
as diffractive charm production. 
The gluon distributions we use are based on 
the model of Ref.\cite{CapellaF2}, hereafter referred to 
as CKMT.
No new parameters are introduced into our calculations.
In this approach there is only one basic Pomeron, which contributes in 
different ways to both ``soft"  and ``hard" mechanisms.
This can be represented in terms of an effective or ``dressed" 
Pomeron with a $Q^2$-dependent intercept.
It was the basis of a very compact parametrization by CKMT 
for the structure function $F_2$ and distributions of 
quarks and gluons in the proton, which was used as an initial 
condition for QCD evolution \cite{CapellaF2}. 
Also diffractive vector meson production ($\rho,\phi,\psi$)
by real and virtual photons \cite{Haakman} could be described
very well by this model without additional parameters.

There have been a variety of investigations on aspects of 
charm production in deep inelastic and diffractive scattering,
in part going beyond the leading order.
In order to achieve a clear and simple discussion of these reactions, 
we confine ourselves to the leading order terms in the cross sections. 
This more phenomenological approach seems reasonable in view of
the large sensitivity of perturbative QCD calculations to the 
heavy quark mass and the present uncertainty in the data. 

In Section $2$ we shall first consider inclusive charm photo- and 
electroproduction on the proton. 
We shall formulate a method of calculation of the 
cross sections of heavy quark production for arbitrary 
values of $Q^2$.
It will be shown that the gluon distribution proposed by CKMT 
gives a very good description of the existing data on charm production.
We therefore proceed in  Section $3$ to calculate charm production in diffractive scattering in the context of the CKMT model 
and compare to other models. 
In Section $4$ we present predictions for inelastic and diffractive 
beauty production by real and virtual photons.

\section{Charm photo- and electroproduction}
In this section we consider the open charm production by 
real or virtual photons on a proton. 
In the past several prescriptions have been proposed to 
incorporate the contribution of charm into $F_2(x,Q^2)$. 
A widely used method \cite{Collins,GRV,MRS,CTEQ} was to 
generate the charm quarks dynamically, starting with no intrinsic 
charm below some threshold $Q_{th}^2 \sim m_c^2$ and to produce 
charm quarks (considered as massless) through QCD evolution. 
However this results into a too large charm contribution in the 
threshold region. 
We therefore choose a different approach.
For moderate $Q^2$, the fact that the mass of the charm quark 
is comparatively large, $m_c \sim 1.5~{\rm GeV} \gg \Lambda_{QCD}$,
makes it possible to apply perturbation theory. 
It was shown in Ref.\cite{Gluck} that therefore more realistic 
predictions for charm production can be obtained
from the photon-gluon fusion diagram of Fig.\,1a. 
This is true in a broad region of $Q^2$.  
Next to leading order (NLO) calculations in QCD perturbation theory 
have been carried out 
\cite{Gluck,Ellis,Smith,Laenen1,Laenen2,Buza,Mangano,Frixione} 
for heavy quark production.
These higher order terms lead to no new qualitative feature 
and can be incorporated by adjusting the parameters of the lowest order calculation \cite{Gluck}.  
In view of this perturbative stability and since there are 
uncertainties in the charm quark mass and the factorization scale, 
we prefer to simply work in the leading order (LO), where the 
calculations are very transparent. 
We shall show that by properly chosing the charm quark mass and 
factorization scale one obtains a good description of the data 
for charm production in DIS.  
For $Q^2\gg m_c^2$, on the other hand, higher order diagrams of the QCD perturbation theory should be resummed \cite{Shifman}.
Since mass effects can then be neglected, the charm quarks
will be produced dynamically through QCD evolution, where the
charmed quark is considered as massless. 
We obtain the charm quark distribution needed as input for this 
evolution through the photon-gluon fusion mechanism.

At moderate $Q^2$, the main contribution for large 
centre-of-mass energies $\sqrt{s}$ is expected to come from 
the gluon fusion process, shown in Fig.\,1a, where a gluon 
from the proton interacts with the photon and produces a charm 
  anti-charm quark pair. 
We will discuss this mechanism first. 
Contributions from ``resolved" charm production, Fig.\,1b, 
are less important and will be dealt with later.

In QCD perturbation theory the cross section for the process 
$ \gamma ^{(*)} p \rightarrow c \bar{c} X$ can be written as a 
convolution of the gluon distribution $g_p(z,\mu_f^2)$ and 
the partonic cross section of the photon-gluon fusion process 
$\hat{\sigma}(  {\gamma g\rightarrow c \bar{c}})$,
\beq
\sigma({\gamma^{*} p\rightarrow c \bar{c} X})=
\int_{z_{min}}^1 \,{d z}\,\hat{\sigma}_{\gamma g
\rightarrow c \bar{c}}(x,z,{\frac{Q^2}
{m_c^2}},m_c^2) \,g_p(z,\mu_f^2)~~.
\label{PPCRS1}
\eeq
The threshold condition for production af a charm-anticharm pair
leads to a lower bound for the fraction $z$ of the proton
momentum carried by the gluon, $z_{min} = a x$, where
$a = 1 + 4 m_c^2/Q^2$ and $x$ is the Bjorken  variable. 
For the partonic cross section of charm production by photon-gluon 
fusion, one finds \cite{Shifman,Witten} with the Hand convention for
the flux
\beq
\hat{\sigma}_{\gamma g\rightarrow c \bar{c}}(x,z,
{\frac{Q^2}{m_c^2}},m_c^2)  = 
\frac{4 \pi^2 \alpha_{\scriptstyle{EM}}}{Q^2 (1-x)} 2 x e_c^2 
\frac{\alpha_s(\mu_F^2)}{2 \pi}
 C\left( \frac{x}{z}, {m_c^2\over Q^2}\right)~~, 
\label{F2cc}
\eeq
where $\mu_f$ is the factorization scale and $e_c$ the charge 
of the charm quark in units of $e$.
In leading order the coefficient function $C$ is
\beqa
C(\zeta, r) & = & \frac{1}{2}[\zeta^2 + (1-\zeta)^2 + 
4\zeta(1-3\zeta) r - 8\zeta^2 r^2] \ln{1+v\over 1 -v}  \nonumber \\
& & + \frac{v}{2} [-1 + 8\zeta(1-\zeta)-4\zeta(1-\zeta)r]~~,
\label{coefch}
\eeqa
with
\begin{equation}
v^2 = 1 - {4 r \zeta \over 1-\zeta} ~~.
\end{equation}
In the limit $Q^2=0$ Eq.(\ref{PPCRS1}) yields the photoproduction
cross section.
The contribution of charm production to the proton structure function
is given by
\beq
F_{2}^{c\bar{c}}(x,Q^2)=\frac {Q^2 (1-x)}{4 \pi^2
 \alpha_{\scriptstyle{EM}}}
\sigma({\gamma^{*}p\rightarrow c \bar{c} X})
\label{defcrs} ~~.
\eeq

For the gluon distribution, a crucial ingredient in the 
calculation of the cross section, we use the CKMT-parametrization
of Ref.\cite{CapellaF2}.
It gives a good description of the HERA data for the proton
structure function $F_2(x,Q^2)$ at all $Q^2$ and 
for diffractive vector meson production \cite{CapellaF2,Haakman}.
We briefly summarize here only the physics background; details can be 
found in the Appendix and in Ref.\cite{CapellaF2}.

Experimental studies of small $x$ DIS at HERA
 \cite{ZEUSF2,H1F2} had shown a fast increase of the proton 
structure function $F_2(x,Q^2)$ as $x\rightarrow 0$, 
which was considered by some authors as an evidence for 
a ``hard" Pomeron. 
This Pomeron has an intercept $\alpha_P(0)$ substantially 
above unity, contrary to the ``soft" Pomeron, which is
observed in high-energy hadronic interactions and 
photoabsorption, which has an intercept only slightly 
above 1:~~$\alpha_P(0)\equiv \Delta + 1\approx 1.08$.
It was argued however by CKMT in Ref.\cite{CapellaF2} that 
this $\alpha_P(0)$ extracted from high-energy 
behaviour of hadronic total cross sections is not the 
actual intercept of the Pomeron itself, but an effective
value that incorporates large effects of Pomeron 
rescattering (multi-Pomeron cuts) in soft processes. 
The actual ``bare" value of the pole intercept 
extracted from the analysis of many features of hadronic 
interactions, taking into account multi-Pomeron processes,
was found to be substantially higher, corresponding to 
$\Delta\approx 0.2 $ \cite{KPTM}.
At large $Q^2$ in DIS the contributions from rescatterings
are much smaller than in hadronic interactions (or $\gamma p$) 
and the ``bare" intercept determines the
behaviour of structure functions at $Q^2 >1$ GeV$^2$.
In this approach there is thus only one Pomeron, which 
contributes to both ``soft" and ``hard" processes.
This effective or ``dressed" Pomeron then has a $Q^2$-dependent 
intercept. 
It was used by CKMT to provide a compact parametrization of 
the structure function $F_2$ and the distributions of quarks
and gluons in the nucleon.
These distributions are used as initial condition for the 
QCD evolution \cite{CapellaF2}. 

The gluon distribution resulting from this model is 
\beq
x g_p(x,Q^2)=C_g x^{-\Delta(Q^2)}(1-x)^{n(Q^2)+3}~~,
\label{PGF}
\eeq
with the effective Pomeron intercept determined by
\beq
\Delta(Q^2)=\Delta(0)\left(1+\frac{d_0 Q^2}{Q^2+d_1}\right)~~.
\label{Pominter}
\eeq
For low $x$ the gluon density in this model exhibits
the $x$ dependence which is characteristic for the
Pomeron exchange.
The behaviour at $x\sim 1$ has been obtained from counting rules 
for the sea quarks which are a factor $1-x$ softer than the gluons.
Finally, the factor $C_g$ follows from the momentum sum rule.
This parametrization is valid up to $Q^2=5$ GeV$^2$. 
We actually use it only up to $2$ GeV$^2$ and for higher 
values QCD evolution is applied where the parametrization 
provides the initial condition.
We take the running coupling constant with $\Lambda=0.20$ GeV and 
four flavours.

The results for charm production are very sensitive to small changes 
in the value of the charm mass. 
We will use charm photoproduction to fix the value of this 
parameter.
In Eqs.(\ref{PPCRS1}) and (\ref{F2cc}) one encounters the 
factorization scale ${\mu_f}$.
The NLO results in Ref.\cite{Gluck} showed that the scale 
$\mu_f^2=4m_c^2$ yields the best stability of the perturbative 
calculations and we will use this now for photoproduction;
other choices will be discussed below in connection with 
electroproduction.
Since this factorization scale is above the value of $Q^2$ 
for which the parametrization is valid, we use the leading 
order QCD evolution equations for three flavours to obtain the gluon distribution at $\mu_f^2$.

We calculated open charm production corresponding to the diagram
of Fig.\,1a with the gluon distribution function as given in 
(\ref{PGF}). 
The results for the total inclusive cross section are shown for four 
different values of the charm mass in Fig.\,2. 
One can see that relatively small changes in the charm mass lead
to large differences in the cross section. 
For a charm quark mass $ m_c = 1.4~ $GeV our curves are close  to
the experimental data and look similar to the results of the NLO
calculation \cite{Gluck,Frixione}.
In the following we shall use this value of the charm quark mass.

For very large energies $\sqrt{s}$ the resolved contribution, 
shown in Fig.\,1b, starts to be important.
In this production mechanism a gluon from the photon 
interacts with a gluon from the proton to produce a charm
anti-charm pair.
To obtain the contribution from the resolved production, we 
therefore need the gluon distribution in the photon, 
$g_{\gamma}(x,Q^2)$.
This distribution is poorly known at present. 
For an estimate of the resolved contribution, we use a distribution
obtained by the same method as for the determination of the gluon
distribution of the proton in Ref.\cite{CapellaF2} and thus 
introduce no new parameters.
At small $x$ factorization takes place in the Regge pole model.
Therefore the gluon distribution in the photon at low $x$ is
proportional to that of the proton; for this proportionality 
factor we find
\beq
e^{\gamma}_p \equiv \lim_{x\rightarrow 0}
\frac{ g_{\gamma}(x,Q^2)}{g_p(x,Q^2)}=\lim_{s\rightarrow \infty}
\frac{\sigma^{tot}_{\gamma p}}{\sigma^{tot}_{pp}} \approx 0.003~~.
\label{FACT}
\eeq
This factorization is approximately valid also when Pomeron cuts 
are taken into account.
The behaviour in the limit $x\rightarrow 1$ is different for the photon
and proton and can be found from counting rules.
So we write $g_{\gamma}(x,Q^2)$ in the form
\beq
xg_{\gamma}(x,Q^2) = e^{\gamma}_p\frac{xg_p(x,Q^2)}{(1-x)^2}= 
e^{\gamma}_pC_gx^{-\Delta (Q^2)}(1-x)^{n(Q^2)+1}~~.
\label{GGF}
\eeq

The contribution of the diagram in Fig.\,1b is shown in Fig.\,2. 
It is negligible at energies up to $\sqrt{s}\sim 10$ GeV, 
but increases with energy faster than the dominant 
photon-gluon fusion contribution of Fig.\,1a. 
This contribution is of the order of $10\%$
already at HERA energies. 
For other models of $g_{\gamma}(x,Q^2)$ this contribution is larger \cite{Frixione}. 
It is also sensitive to the value of the factorization scale and 
increases for smaller values of this scale.

We now turn to electroproduction. 
First we will consider the factorization scale $4 m_c^2$ as in
photoproduction.
The production of charm quarks give a large contribution to 
the DIS structure function $F_2(x,Q^2)$ at small $x$ and large
$Q^2\gg m_c^2$, where they can be considered in the framework 
of QCD evolution equation as massless. 
However at $Q^2 \sim 10~$GeV$^2$ or less the charm mass is very 
important and the charm quarks cannot be treated in the same way 
as the light quarks \cite{Gluck}.  
We therefore deal with electroproduction in two different ways, 
depending on the value of $Q^2$. 
For values of $Q^2$ less than a certain value 
${\overline {Q^2}}$, we use the perturbative approach
for the gluon fusion process with massive charm quarks 
as given by Eq.(\ref{PPCRS1}).  
To obtain the gluon distribution at the factorization scale 
$4 m_c^2$, we proceed as in the photoproduction case;
we start the QCD evolution for three massless
flavours from $Q_0^2=2$ GeV$^2$.
With this distribution, the charm contribution is directly 
obtained from Eq.(\ref{F2cc}). 
For virtualities larger than ${\overline {Q^2}}$ the charm 
quarks are produced through massless QCD evolution with
four flavours. 
For the input distributions, we use the light quark
distributions as obtained above from three flavour
massless QCD evolution up to ${\overline {Q^2}}$.
The charm input distribution is generated by photon-gluon 
fusion at ${\overline {Q^2}}$.

We determine the transition value ${\overline {Q^2}}$  
from the perturbative approach with massive charmed quarks 
to the massless QCD evolution by demanding that this procedure 
creates an $F^{c\bar{c}}_2$ below and above ${\overline {Q^2}}$ 
with a smooth derivative with respect to $Q^2$ at this point.
We found that the values of ${\overline {Q^2}}$ yielding
a smooth transition in the region of small $x$ vary from $30$ 
GeV$^2$ to $100$ GeV$^2$.
In Fig.\,3a we show the charm contribution to the proton structure
function, $F_2(x,Q^2)$, as given by Eq.(\ref{F2cc}) as a function of
the virtuality $Q^2$ at different values of $x$;
the transition from massive to massless treatment of
the charm quarks is made at $50$ GeV$^2$.
We also show the logarithmic derivatives in Fig.\,3b to see their 
discontinuities at ${\overline {Q^2}}$.
The change of $\Lambda$ in the running coupling constant 
at the charm threshold has been taken into account. 
It follows from Figs.\,3a and 3b that for very small $x$ ($x\le 10^{-3}$)
the transition from one regime to the other is very smooth 
and for $Q^2~>~50~ $GeV$^2$ charmed quarks can safely be 
considered as massless in the QCD evolution equations. 
This is in agreement with results obtained in Ref.\cite{Laenen2} 
and also with Ref.\cite{Gluck}, where it has been stated that for 
$W^2=Q^2(1-x)/x\leq 10^6$ GeV$^2$ the gluon fusion model should 
be applied.
For $x\sim 0.1$ the mass effects are important up
to much larger values of $Q^2$. 
Thus, from Fig.\,3a we can conclude that for the small $x$ 
and $Q^2 \lesssim 500$ GeV$^2$ the difference between the
prediction of the gluon-photon fusion diagram and its QCD-evolved 
contribution is small.

So far, we have used the factorization scale $4 m_c^2$.
Another natural candidate for electroproduction is $4m_c^2+Q^2$. 
To examine this possibility, we show in Fig.\,3c and 3d the analogous 
results for this scale. 
The features are quite different.
Now for large $x$ the discontinuity is small, but for small $x$ 
it is large, just the reverse of what one finds for
the constant factorization scale.
This suggests that one could use a different factorization scale
and/or different transition value ${\overline {Q^2}}$ for
different kinematical regimes to obtain a smooth transition from
massive to massless quarks.

Finally, in Fig.\,4 we make a comparison of our predictions for 
charm photo- and electroproduction with the data.
We show results as a function of $W = \sqrt{s}$ for the 
factorization scales $4 m_c^2$ and $4 m_c^2 + Q^2$;
for the latter we only show predictions at high $Q^2$ 
where the difference to the fixed scale is large.
Note that the cross sections for the different values of 
$Q^2$ in Fig.\,4 are rescaled. 
The photoproduction data at high energies are from ZEUS and H1 
\cite{ZEUS1,H11}; the other experiments are listed in 
Ref.\cite{ZEUS1}. 
The low energy data for electroproduction are from 
Ref.\cite{NMC,Landshoff} and the high energy data 
from H1 \cite{H1}.
The latter data correspond to $Q^2$ values slightly different
from those of the low energy data and of our calculations.
We see that all data are rather well described. 
The high $Q^2$ data indicate a preference for the fixed
factorization scale of $4 m_c^2$ : the charm production at 
different (but not very large) $Q^2$ depends on the gluon 
distribution at a fixed scale.
Therefore we will use this factorization scale also for 
diffractive charm production. 

The deviation of the H1 results at $Q^2=12$ GeV$^2$ from our
theoretical prediction is partly explained by the fact that for 
our prediction we took the same $Q^2$ as for the low energy data, 
{\em i.e.}~$Q^2=13.9$ GeV$^2$.
It means that our predictions are here slightly too small.
Nevertheless, the discrepancy remains significant and is
difficult to explain in view of the fact that for other 
values of $Q^2$ the agreement between theory and experiment is 
reasonable and that the cross sections have only a weak $Q^2$ 
dependence in this region.
Except for these H1 points, our comparison confirms the 
photon-gluon fusion mechanism in combination with the CKMT gluon 
distribution at the factorization scale $4m_c^2$. 

In comparing our results to the data, it should be noted that the 
``resolved" diagrams of the type shown in Fig.\,1b
can contribute also for highly virtual photons. 
Our estimates indicate that this contribution at 
present energies is rather small --- it is less than $10\%$
of the main diagram of Fig.\,1a. 
However in the same way as for real photons the contribution 
of Fig.\,1b increases with energy faster than for Fig.\,1a.
Thus it will dominate at superhigh energies where it 
corresponds to charm production in the central rapidity region.

\section{Hard diffractive scattering and charm production}
In deep inelastic scattering, there are events where 
particles are produced only in the fragmentation regions of the 
initial particles.
We consider here the case that the proton remains intact, the so
called ``single diffractive scattering". 
This corresponds to a large ``rapidity gap" between the diffractively
produced states (with invariant mass $M_X$) and the recoil
proton. 
This observation is readily explained in the Reggeon 
theory through the exchange of the Pomeron (Fig.\,5a). 

In this section, we discuss the 
diffractive cross section in the Pomeron exchange model 
\cite{Ingelman} and specifically consider diffractive charm 
production. 
As in the total inclusive charm production discussed above, 
gluon fusion is again the dominant production mechanism for charm. 
Therefore, diffractive charm production is sensitive to the
gluon distribution in the Pomeron, which is believed to consist 
mainly of gluons. 
This should be compared to the analogous inclusive process,
where mainly the gluon component of the {\it proton} is probed.
We will use the CKMT approach to obtain the parton distributions
in the Pomeron and compare its results to two other models for the 
Pomeron parton distributions.

The differential production cross section can be written as 
\beq
\frac {d^4\sigma_{\rm DIF}}{dx dQ^2 dx_P dt}=\frac{4\pi \alpha_{EM}^2}
{x Q^4}\left\{ 1-y+\frac{2y^2}{2\left[1+R_{\rm DIF}(x,Q^2,x_P,t)\right]}
\right\} F_2^D(x,Q^2,x_P,t)
\label{DIFcrs}~~,
\eeq
where we have introduced a diffractive structure function 
$F_2^D(x,Q^2,x_P,t)$.
Here $x, y$ and $Q^2$ are the usual DIS variables and $t$ is the
squared momentum transferred to the proton, $t\equiv k^2=(p-p')^2$.  
The variable $x_P$ is defined by
\beq
x_P=\frac{q \cdot k}{q\cdot p}\simeq \frac{M_X^2+Q^2}{W^2+Q^2}~~,
\label{xpom}
\eeq
with $M_X^2=(q+k)^2$ the squared invariant mass of the diffractively
produced particles and $W^2=s=(q + p)^2$, the squared CMS energy 
of the photon-proton system.
For the mechanism shown in Fig.\,5a, the variable $x_P$ can be 
interpreted as the fraction of the proton momentum carried by
the Pomeron.
The function $R_{\rm DIF}$ in Eq.(\ref{DIFcrs}) is the ratio of
the longitudinal to transverse part of the cross section.
Integrating $F_2^D(x,Q^2,x_P,t)$ over $x_P$ and $t$ we obtain 
the diffractive contribution to the total deep inelastic 
structure function $F_2(x,Q^2)$. 

For the exchange of the Pomeron pole, the diffractive structure 
function $F_2^D(x,Q^2,x_P,t)$ can be factorized into two parts,
\beq
F_2^{D}(x,Q^2,x_P,t)=f(x_P,t)F_P(\beta,Q^2,t)~~,
\label{F2d}
\eeq
where the variable $\beta=Q^2/(M_X^2+Q^2) = {x}/{x_P}$ plays 
the same role as the Bjorken variable $x$ has in DIS:
the Pomeron momentum fraction carried by the partons in the Pomeron.
The first factor represents the Pomeron flux from the 
proton and can be written in the form
\beq
f(x_P,t)=
\frac{\left( g_{pp}^P(t)\right)^{2}}{16\pi}
x_P^{1-2\alpha_P(t)}~~,
\label{PomFlow}
\eeq
where $g_{pp}^P$ denotes the Pomeron-proton coupling.
The value of $\alpha_P(t)$ in this flux factor 
should be taken at some effective virtuality 
scale, $Q_{eff}^2$, since in the model of Ref.\cite{CapellaF2}
one has
\beq
\alpha_P(t) = 1 + \Delta(Q_{eff}^2) + \alpha_P' t~~,
\eeq
where $\Delta$ is given by Eq.(\ref{Pominter}). 
The scale $Q_{eff}^2$ is 
{\it a priori} not known, but it was argued in Ref.\cite{CapellaFP1} 
that it should be low, since a hard scale from the top part of 
Fig.\,5b does not get through to the lower part of the diagram. 
From theoretical point of view, values of $\Delta(Q_{eff}^2) = 0.13$ 
to $\Delta(Q_{eff}^2)= 0.24$ are possible, 
corresponding to the effective Pomeron intercept without
eikonal-type corrections and the ``bare" value, respectively.
Both extremes are not excluded by experimental evidence
\cite{ZEUSD,H1D}.
The Pomeron slope has its usual value $\alpha_P'=0.25$ GeV $^{-2}$.
The second function, the Pomeron structure function $F_P$, is 
proportional to the virtual photon-Pomeron cross 
section. 
It was emphasized in Ref.\cite{CapellaFP1} that other definitions of 
the ``flux factor" are possible, {\it e.g.} differing 
from the one above by a constant. 
Therefore the normalization of the Pomeron structure function 
depends on the particular choice of the flux factor. 
For large values of $M_X$ or small values of $\beta$, the 
Mueller generalization of the optical theorem for inclusive
cross sections can be used to represent the cross section 
given by Eq.(\ref{DIFcrs}) in terms of the
triple-Reggeon diagram shown in Fig.\,5b. 
The Reggeon exchange in the upper part of the diagram is dominated
by the Pomeron and $f$ Regge poles.
Since each term factorizes into a Reggeon propagator and a vertex, 
it is possible to obtain the structure function of the Pomeron 
from that of the deuteron \cite{CapellaFP1}.

We will below consider the diffractive production involving 
light quarks and heavy quarks separately,
{\em i.e.} we write the Pomeron structure function as
\beq
F_P = F_P^{0} + F_P^{c\bar{c}}~~.
\eeq
For the contribution of light quarks to the 
structure function, we adopt the form for the deuteron 
structure function of Ref.\cite{CapellaF2}, but with the 
couplings of the exchanged Reggeon to the deuteron replaced 
by the couplings to the Pomeron,
\beq
F_P^0(\beta,Q^2,t)= e^{f}_{d}C_{f}^d(\beta,Q^2) \beta^{1-\alpha_f}
(1-\beta)^{n(Q^2)-2}+ e^{P}_{d}
C_P^d(Q^2) \beta^{-\Delta(Q^2)}(1-\beta)^{n(Q^2)+2}~~,
\label{PQP}
\eeq
with the ratios of the coupling constants
\beq
e^{k}_{d}=\frac{r_{PP}^k(t)}{g_{dd}^k (0)}~~,~~
\eeq
where $r_{PP}^k$ and $g_{dd}^k$ are 
the couplings of the Pomeron $(k=P)$ or the leading $f$ Regge 
pole $(k=f)$ to the Pomeron and to the deuteron, respectively.
The values for the ratios $e^{k}_{d}$ can be estimated from 
soft diffraction data.
In our calculations we will use $e^{P}_{d}=e^{f}_{d}=0.07$, which
was shown in Ref.\cite{CapellaFP2} to give a good description of the
data on the diffractive structure function $F_2^D$.
The behaviour of the structure function for $\beta\rightarrow 1$ 
is determined by the exponent $n(Q^2)$ which was obtained for the
deuteron in Ref.\cite{CapellaF2} by using counting rules.
For the Pomeron the exponent has to be adjusted accordingly since 
there is one parton spectator less.

From experiment it follows that the triple Reggeon couplings
are only weakly dependent on $t$ and that for the Pomeron and the $f$ 
Regge pole exchange this dependence is approximately equal.
It can then be incorporated in that of
the flux factor in Eq.(\ref{F2d}), {\it i.e.} in $g_{pp}^P$,
which is then parametrized as $g_{pp}^{P}(t)=g_{pp}^{P}(0) \exp(Ct)$
with $C=2.2$ GeV$^{-2}$ and $g_{pp}^P(0)^2=23$ mb.
With this parametrization, the structure function of the Pomeron 
(or its parton distributions) do not have the $t$-dependence
any more. 
The values of the parameters mentioned above have been all
taken from Ref.\cite{CapellaFP1}. 
We again use the parametrization at an initial
scale of $Q^2_0 = 2$ GeV$^2$ as starting point for QCD evolution
to the desired $Q^2$ value.
The contribution of the light quarks to the Pomeron structure 
function, $F_P$, is directly related to the corresponding
singlet quark distribution of the Pomeron,
\beq
F_P^0(\beta, Q^2)= \frac{2}{9} \Sigma_P(\beta, Q^2)~~,
\eeq
with  
\beq
\Sigma_P(\beta, Q^2)=\sum_{i=u,d,s}\left[ \beta q^i_P(\beta,Q^2)+ 
\beta {\bar q}^i_P(\beta,Q^2)\right]~~. 
\eeq
This singlet distribution at $Q_0^2$ is shown in Fig.\,6a.

We now discuss the charmed quark contribution to the diffractive
Pomeron structure function, which is related to the analogous 
charm quark distribution,
\beq
F_P^{c\bar{c}}(\beta,Q^2)= \frac{4}{9} \left[ \beta c_P(\beta,Q^2)+ 
\beta {\bar c}_P(\beta,Q^2)\right]~~. 
\eeq
Using photon-gluon fusion as the leading mechanism, the diffractive 
charm production can be calculated in the same way as for 
charm production in DIS in Section $2$. 
This mechanism for diffractive production is shown in Fig.\,7a. 
Thus while charm production in DIS was a probe of the gluon 
distribution in the proton, the diffractive production
of charm now probes the gluon distribution in the Pomeron, 
which is our main point of interest here.
In addition to the ``sea" contribution in Fig.\,7a,
there is also a ``valence" contribution from the Pomeron
structure function. 
A typical example of such a contribution can be seen in Fig.\,7b, 
which has been studied in Ref.\cite{Nikolaev}. 
The role of these additional mechanisms, which require further model assumptions, is not clear yet and had not been discussed in other 
papers on diffractive charm production.
While the ``sea" component is concentrated mostly at small values of 
$\beta$, it can be shown that the ``valence" contribution 
has its maximum at $\beta\sim 0.5$ in Fig.\,7b.
Its value is smaller by a factor of order $1/m_c^2$.
In this region the ``valence" contribution can therefore become 
comparable to the ``sea" component.

In the CKMT approach, the gluon distribution of the Pomeron can for
low $\beta$ be obtained from that of the proton (or deuteron), 
analogously to the Pomeron quark distribution discussed above. 
For a large parton momentum fraction, however, the analogy 
breaks down.
The gluon distribution in the nucleon is determined at large 
$x$ by the gluon radiation originating from the quarks; in the
Pomeron, which is believed to be mainly composed of gluons,
they are {\it a priori} present. 
In Ref.\cite{CapellaFP2} therefore the large $\beta$ behaviour 
was modified according to 
\beq
\beta g_P(\beta, Q^2)=e^{P}_{d} C_g\beta^{-\Delta(Q^2)}
(1-\beta)^{n_g}~~,
\label{GDP}
\eeq
where $n_g$ is a free parameter. 
The distributions we use are thus singular at $\beta = 0$ due to the $\beta^{-\Delta}$ dependence dictated by the Pomeron exchange. 
The observed $Q^2$ dependence of the data indicate that the gluon 
distribution should be rather hard and thus $n_g$ negative, {\it i.e.}
between $0$ and $-1$ to yield a normalizable distribution.
We use in the following two negative values of $n_g$, and therefore 
our gluon distribution is also singular for $\beta = 1$. 
This can be seen in Fig.6b, where we show our distributions for a 
scale of $Q^2_0 = 2$ GeV$^2$. 
For the actual calculation of diffractive charm production, 
we use QCD evolution to obtain the gluon
distribution at the factorization scale of $4 m_c^2$.

Before showing our results for the diffractive production of light
as well as charmed quarks, we first discuss the models of Refs. \cite{Gehrman,Golec,H1P}, where a similar description of the
reaction involving the Pomeron structure function was used: 
model I of Gehrmann and Stirling (GS) \cite{Gehrman} and 
model 3 of Golec Biernat and Kwiecinski (GK) \cite{Golec}. 

GK take the following singlet quark and gluon distributions, 
respectively, at $Q^2_0=4$ GeV$^2$:
\beq
\Sigma_P(\beta)=0.069~ K~ \beta^{0.44}(1-\beta)^{0.60}~~,~~
\beta g_P(\beta)=1.16~ K~ \beta^5~~. \label{GKpar}
\eeq  
The constant $K$ relates the measured $t$-integrated diffractive
structure function to the Pomeron structure function:
\beq
\tilde{F}_2^{D}(x,Q^2,x_P)\equiv\int dt~F_2^D(x,Q^2,x_P,t)=
K \left(\frac{1}{x_P}\right)^{\Delta} F_P(\beta,Q^2)~~.
\eeq
The value for $K$ depends on $x_P$ and is approximate 10
for the current experiments.
The distributions used by GS at $Q_0^2=2$ GeV$^2$ are:
\beq
\Sigma_P(\beta)=1.02~ \beta(1-\beta)~~,~~
\beta g_P(\beta)=4.92~ \beta(1-\beta)~~. \label{GSpar}
\eeq
In order to compare their quark and gluon distributions to ours,
one has to take into account that different normalizations of the 
assumed Pomeron flux, $f(x_P,t)$, are used and they have to be
converted to our conventions.

In Fig.\,6a and 6b we compare the different model distributions 
for quarks and gluons in the Pomeron, all normalized according 
to our definition of the flux factor.
We show the distributions as originally parametrized on their
respective initial scales, $Q^2_0$.
The quark singlet distribution, $\Sigma_P$, looks similar in the
three models. 
Except for the behaviour of the CKMT parametrization at 
very low $\beta$, the singlet quark distributions show 
roughly the same qualitative features for all models, because 
they were fitted to diffractive production data for $\beta \ge 0.065$. 
Some differences with the CKMT model are due to the fact
that the other models fit to all the diffractive data and thus also
simulate doubly diffractive data with their fit, 
which account for about $30 \%$ of the cross section.

In contrast to the quark distributions the gluon 
distributions of the three models are dramatically 
different: the predicted distributions have different shapes.
The GK parametrization is the ``hardest" of the
three types of distributions, remaining essentially zero for low
beta and having its main strength near $\beta = 1$. 
The distribution of GS is symmetric, vanishing at $\beta = 0$ 
and $\beta = 1$, and thus is peaked in the middle. 
The gluons in the CKMT model are distributed with nearly constant 
density for intermediate $\beta$ values; their distribution diverges
at both ends.
It becomes steeper at $\beta = 1$ the closer $n_g$ gets near $-1$. 
The very large values of $g_P(\beta)$ for intermediate values of 
$\beta$ in the GS and GK models are difficult to reconcile 
with the cross sections for jet production in diffractive hadronic 
interactions \cite{Goulianos}. 
The most direct test of all models is provided by the 
diffractive production of heavy quarks and we will therefore 
now look at diffractive charm production in DIS.

For the calculations, the gluon distributions are used
at the common factorization scale of $4 m_c^2$. 
They are shown in Fig.\,6d and can be seen to have rather different 
qualitative features than in Fig.\,6b. 
The QCD evolution from the initial scale $Q_0^2$ to the higher scale
shifts the gluon distributions towards lower momentum fractions. 
This can be most clearly seen for the GS model.
All gluon distributions are now singular at $\beta = 0$.
For GS and GK this is due to the evolution, while CKMT already
starts out singular at the origin due to the 
$\beta^{-\Delta}$ behaviour, which is only somewhat 
modified by QCD evolution.
The opposite effect can be seen at $\beta = 1$: the originally
singular distribution in the CKMT model becomes zero. 
For completeness, we also show an
example for the effect of evolution on the light quark 
distribution in Fig.\,6c.
For GS and GK, the magnitude of the distribution increases for
moderate values of $\beta$.
Furthermore, again a slight shift towards small $\beta$ occurs.
For CKMT, the QCD evolution makes the density becomes more flat
and the change in magnitude is such that the distribution is lower 
than the others at intermediate $\beta$.

One has to be careful when considering the energy-momentum sum rule,
{\em i.e.} the integral over $\beta$ of the sum of the gluon and quark 
distributions times $\beta$.
As was pointed out before, the choice of the Pomeron flux factor
fixes the normalization of the Pomeron structure function and
 therefore the value of the energy-momentum sum rule. 
This sum rule needs not to be equal to 1 as was assumed in several 
publications. 
However, the value of this energy-momentum sum is preserved under
QCD evolution.
``virtual particle".
This explains the changes in magnitude discussed above: 
in response to an overall growth in the quark density 
the gluon density decreases.
At the factorization scale, the total gluon content of the
Pomeron predicted by CMKT is $60\%$ and about $80\%$ for 
the other models. 

The contribution of charm quarks to the Pomeron structure function, 
produced by the gluon-photon fusion mechanism is shown in Fig.\,8
for the three gluonic distributions discussed above and for 
different values of $Q^2$. 
The results are obtained by carrying out a convolution analogous to electroproduction. 
To obtain $F_P^{c\bar{c}}$ at a given momentum fraction $\beta$ 
carried by the struck charmed quark, we have to integrate over
gluon momentum fractions starting from a $Q^2$ dependent cut-off 
value that follows from the lower limit in Eq.(\ref{PPCRS1}).
Both the CKMT and GS result increase continously as $\beta$ 
decreases from this cut-off.
The GK result has a maximum about halfway which originates
from its relatively hard gluon distribution.
As the virtuality of the photon increases, the cut-off moves to 
larger values, but the general shape of the curves doesn't change. 
For a $Q^2$ value as high as $500$ GeV$^2$, the factorization 
scale we have chosen may not be appropriate anymore and a $Q^2$ 
dependent value might be better.
In general, the shown predictions are sufficiently different 
such that future experiments for charm production in hard
diffractive scattering should be able to discriminate between
them.  

To see how significant the charm production contributes to 
the total diffractive cross section, we also show in Fig.\,8 
the full $F_P$.
Adding these charm quark contributions results in a (small) 
$Q^2$ dependent violation of the sum rule discussed above.
The fraction of the total structure function provided by the charm 
is in all cases most important for small $\beta$ and 
increases with $Q^2$. 
A measurement of the total diffractive cross section is of 
course a test of the different parton density input.  
contributions, 

The exponent $n_g$, which determines the ``hardness" of the 
CKMT gluon distribution, is a free parameter. 
To see how sensitive the results are to its value, we compare 
in Fig.\,9 results for $F_P$ and $F_P^{c\bar{c}}$ for 
$n_g=-0.5$ and $-0.9$ at the same $Q^2$ values as in Fig.\,8. 
The charm part, $F_P^{c\bar{c}}$, is most directly sensitive 
to the gluon distribution.
The shape of the curves does not change much. 
However, as $F_P^{c\bar{c}}$ receives its contribution from 
gluons with high momentum fraction $\beta$ due to the presence 
of the cut-off in the convolution Eq.(\ref{PPCRS1}),
the harder distribution, $n_g = -0.9$, yields a significantly 
larger prediction. 
This is most pronounced at high $\beta$ and
similar for all values of $Q^2$.
For the total $F_P$, the more singular gluon distribution results in
an increase, in particular at larger $Q^2$. 
The exponent  $n_g$ enters on the one hand through the charm
 contribution discussed above. 
On the other hand, it also influences the light quark 
contribution through the QCD evolution.
The curves clearly show that the hardness of the
gluon distribution is relevant for the total diffractive 
cross section.

It is interesting to see the contribution of the diffractive
production to the total DIS structure function, $F_2(x, Q^2)$. 
To compare to the results in Section 2, we convert this into the
virtual photon cross section, $\sigma(W, Q^2)$.
For this purpose, we must integrate the diffractive contribution in 
Eq.(\ref{F2d}) over $x_P$ and $t$. 
For $x_P$ we integrate up to $0.1$, which defines our lower limit
for the rapidity gap.
We use for the Pomeron intercept in the flux factor a value
$\alpha_P(0)=1.13$ as in Ref.\cite{CapellaFP2}.
We show in Fig.\,10 these diffractive contributions at the same 
$Q^2$ values as for the total in Fig.\,4. 
They increase rapidly from their threshold and then reach 
an approximately powerlike $W$-dependence. 
Comparison of Figs.\,4 and 10 shows that the diffractive production 
is for large energies an order of magnitude smaller than 
the total. 
For large energies, the result for the harder gluon 
distribution is about a factor 1.5 higher.

 \section{Beauty Production}
We now repeat some of the above calculations for the
production of beauty quarks. 
Due to the larger mass, this of course involves a different 
factorization scale and thus probes the gluon distribution at 
a much higher scale; we take it as $4 m_b^2$.
The cross section can be calculated using the same photon-gluon 
fusion mechanism and the approach discussed above. 
The only difference is the mass of the b-quark, which we take as
$m_b=4.7~$ GeV. 
It can be found using the mass of the charm quark from the 
following relation
\beq
m_b - m_c = \bar{m}_B -\bar{m}_D~~,
\label{bMASS}
\eeq
which is valid in QCD up to small corrections $\sim 1/m_Q$ 
\cite{Bigi}. 
The quantity $\bar{m}_B\equiv (m_B+3m_{B^{\ast}}/4)$ in 
Eq.(\ref{bMASS}) denotes the center of gravity value for the lowest 
mesonic state (the same holds for $\bar{m}_D$).

Predictions for the cross section for beauty production by real
and virtual photons are shown in Fig.\,11. 
The threshold is higher than for charm production,
but the general shape of the cross section is similar.
The magnitude is typically two orders of magnitude lower, 
which is due to the larger mass and the smaller charge.
The contribution of diffractive beauty production to 
$F_P(\beta, Q^2)$ is shown in Fig.\,12. 
The $\beta$ value for which the result becomes zero has decreased 
due to the larger quark mass.
The shape of the curves, their dependence on $Q^2$ and the
hardness of the gluon function are similar as for charm 
production, Fig.\,9.
The contribution to the structure function $F_2(x, Q^2)$ is shown
in Fig.\,13.
The threshold energy is now higher, but for the rest a similar
behaviour to charm production can be seen: as in Fig.\,10, 
the diffractive contribution is again an order of magnitude 
smaller than the total DIS beauty production and the sensitivity to
the hardness of the gluon distribution looks the same.
In general, these cross sections are much smaller than in the 
case of charm and large statistics is needed to observe them 
experimentally.
\section{Summary}
 
The main goal of this paper was to probe the gluon distribution
in the Pomeron by means of diffractive heavy quark (charm, beauty) 
production. 
Given the present uncertainty in the data as well as the high 
sensitivity of perturbative QCD calculations to the heavy quark mass, 
we have here pursued a simple phenomenological approach to obtain 
a description of the main features of heavy quark production.
First we showed that the data for total open charm photo- and 
electroproduction could be well described by using the gluon 
distribution of the proton predicted by the CKMT model, 
together with the photon-gluon fusion mechanism. 
From this comparison to the data we extracted a value of 
$m_c = 1.4$ GeV for the charm quark mass parameter and
confirmed the finding of Ref.\cite{Gluck} that $4 m_c^2$ is a
good factorization scale. 
Having established this basis, we proceeded to diffractive 
charm production.
We obtained results with gluon distributions of different 
models and compared their contribution to the total diffractive
structure function of the Pomeron. 
For the CKMT model, we investigated how the results depend on
the high momentum behaviour of the gluon distribution. 
A similar study was made for beauty production.

In conclusion we found that different models for the gluonic content 
of the Pomeron lead to sizeable differences for diffractive charm 
production which experiments, such as the future high statistics 
experiments at HERA \cite{Eichler}, should be able to distinguish. 
For beauty production, similar features were found, but the 
cross sections are much smaller and thus more difficult to measure.


\vspace*{1cm}
\centerline{{\bf  ACKNOWLEDGEMENTS}}

The work of L.H. and J.K. is part of the research program
of the Foundation for Fundamental
Research of Matter (FOM) and the National Organization for
Scientific Research (NWO). The collaboration with ITEP
is supported in part by a grant from NWO
and by grant 93-79 of INTAS. A.K. also acknowledges support
from grant 96-02-19184 of RFFI.

\section*{Appendix}
For the QCD evolution equation one needs the quark distributions
at an initial scale $Q_0^2$. 
Up to $Q^2=5$ GeV$^2$ the total valence and sea quark distribution
in LO can directly be extracted from $F_2(x,Q^2)$ as 
parametrized by the CKMT model in Ref.\cite{CapellaF2}:
\beqa
F_2^{val}(x,Q^2)&\equiv&\frac{4}{9}u^v(x,Q^2)+\frac{1}{9}d^v(x,Q^2)=
C_{f}(x,Q^2) x^{1-\alpha_f}(1-x)^{n(Q^2)}~~,\nonumber\\
F_2^{sea}(x,Q^2)&\equiv&\frac{8}{9}u^s(x,Q^2)+
\frac{2}{9}d^s(x,Q^2)+\frac{2}{9}s^s(x,Q^2)=
C_{P}(Q^2)x^{-\Delta(Q^2)}(1-x)^{n(Q^2)+4}~~,
\label{PQF}
\eeqa
with the functions
\beq
\begin{array}{lcl}
C_{f}(x,Q^2)=B(x)\left(\frac{Q^2}{Q^2+b}\right)^{\alpha_R} & ~,~&
C_{P}(Q^2)=A\left(\frac{Q^2}{Q^2+a}\right)^{1+\Delta(Q^2)},\\
\Delta(Q^2)=\Delta(0)\left(1+\frac{d_0 Q^2}{Q^2+d_1}\right) &~,~ &
n(Q^2)=\frac{3}{2}\left(1+\frac{Q^2}{Q^2+c}\right)~~.
\end{array}
\label{parameters}
\eeq
This parametrization has been constructed in such a way that
for $x\sim 1$ it is in accordance with the dual parton model at
low $Q^2$ and with dimensional counting rules at very large $Q^2$.
The low $x$ behaviour is readily explained in terms of
Reggeon exchanges; the secondary Regge trajectory with intercept 
$\alpha_f$, corresponding to ($f,A_2$)-exchanges, 
determines the small $x$ distributions
for the valence quarks, while the ``effective" Pomeron exchange 
determines them for the sea quarks (and the gluon).
The exponents in $C_f$ and $C_P$ are chosen such that the 
photolimit ($Q^2,x \rightarrow 0$) is finite.
  
The function $B$ was written as the sum of the $u$ and $d$ quark
contribution, $B=B_u+B_d$, and is different for the 
proton and the deuteron:
\beq
\begin{array}{lcll}
B_u=\frac{4}{9}C_u &,& B_d=\frac{1}{9}C_d(1-x) & 
{~~\rm for~the~ proton}~~,
\\   
B_u=\frac{2}{9}\left[C_u+C_d(1-x)\right] &,& 
B_d=\frac{1}{18}\left[C_u+C_d(1-x)\right] & 
{~~\rm for~the~deuteron}~~,
\end{array}
\eeq
where the constants $C_u$ and $C_d$ were obtained from the condition 
that the  integrals of the valence distributions over $x$ should 
give the correct number of valence up and down quarks in the proton. 
Here the constants for the deuteron are related to those 
of the proton through isospin symmetry.
For the sea quarks the simple assumption is used that
$u^{s}=d^{s}$ and $s^{s}=0.5\,u^{s}$, in reasonable 
accordance with the results from $\nu N$ interactions.

Also the gluon distribution for low $x$ is determined by the
Pomeron exchange and is thus proportional to the sea quark
distributions.
However, since sea quarks are produced mostly by gluons,
the distribution of gluons will be harder than the one of
the sea quarks.
This leads to the gluon distribution in Eq.(\ref{PGF}):
\beq
x g (x,Q^2)= G \frac{F_2^{sea}(x,Q^2)}{1-x}=C_{g}(x,Q^2)
x^{-\Delta(Q^2)}(1-x)^{n(Q^2)+3}~~,
\eeq
where the proportionality factor $G$ was obtained from
the momentum sum rule at $Q^2=2$ GeV$^2$.

The parameter $d_0$ in the effective Pomeron intercept
was originally \cite{CapellaF2} put equal to $2$ in order 
to have  for the bare Pomeron a value of $\Delta_{bare}$ 
that is 3 times larger than $\Delta(0)$, in agreement with 
analyses of hadronic interactions\cite{KPTM}. 
The seven remaining free parameters of the model were fitted in Ref.\cite{CapellaF2} to the data on total $\gamma p$ interaction 
cross section and to the proton structure function data of NMC.
They are given by
\beq
\begin{array}{ccccccc}
a=0.2631~~ &,& ~~b=0.6452 &,& ~~c=3.5489 &,& ~~d_1=1.1170~~~, \\
A=0.1502~~ &,& ~~\Delta(0)=0.07684 &,& ~~\alpha_f=0.4150 &.&
\end{array}
\eeq
Here the parameters $a,b,c$ and $d_1$ are given in units of
GeV$^2$.
With these values for the free parameters one finds explicitly 
for the coefficients:
\beq
C_u=2.714~~,~~ C_d=1.618~~,~~G=12.63~~.
\eeq
However, using new HERA data in the region of $Q^2 < 2 $ GeV$^2$
\cite{ZEUSNew,H1New} it is possible to determine the parameters $d_0$
and $d_1$ with better accuracy: $d_0=2.2~,~~d_1=0.6$ GeV$^2$.

\small{

}

\newpage

\section*{\bf \large Figure Captions}
\begin{figures}{}
\item{{\bf Fig.\,1}} 
{\bf a.} Photon-gluon fusion diagram.\\
{\bf b.} Charm production by the gluon-gluon fusion mechanism 
(``resolved" contribution). 
\item{{\bf Fig.\,2}} 
Charm photoproduction cross section predicted by the CKMT model.
Solid lines: direct contribution for different values of the 
charm mass; from top to bottom $m_c= 1.3 , 1.4 , 1.5 , 1.6$ GeV. 
Lower dashed line: resolved contribution with $m_c=1.4$ GeV 
obtained with the gluon distribution in the photon according to the 
CKMT model.
Upper dashed line: the sum of the direct and resolved contribution
($m_c=1.4$GeV).
\item{{\bf Fig.\,3}} 
{\bf a.} Charm contribution to the proton structure
function $F_2(x,Q^2)$ in the CKMT approach
as function of $Q^2$ for different $x$; from top to bottom: 
$x= 0.0001, 0.001, 0.01, 0.1$.
Solid lines are obtained by using the gluon fusion process 
below $Q^2=50$ GeV$^2$ and massless QCD evolution for
$Q^2>50$ GeV$^2$.
Dashed lines: charm contribution by using the gluon
fusion model for all $Q^2$.
Factorization scale is $\mu_f^2=4 m_c^2$.\\
{\bf b.} Logarithmic derivative of $F_2^{c\bar{c}}$ obtained
from Fig.\,3a for $x=0.1$ (dot-dashed), $x=0.01$ (dotted), $x=0.001$
(dashed) and $x=0.0001$ (solid line) \\
{\bf c.} Same figure as {\bf a.}, but with factorization
scale $\mu_f^2=4 m_c^2+Q^2$.\\
{\bf d.} Same figure as {\bf b.}, but with factorization
scale $\mu_f^2=4 m_c^2+Q^2$.
\item{{\bf Fig.\,4}} 
Comparison of the charm photo-and electroproduction cross 
sections predicted in this paper with experiment for different 
$Q^2$; from top to bottom $Q^2=0, 1.39, 2.47, 4.39, 7.81, 13.9, 
24.7, 43.9, 78.1$ GeV$^2$.
Solid lines: $\mu_f^2=4 m_c^2$.
Dashed lines: $\mu_f^2=4 m_c^2+Q^2$.
All curves are rescaled with powers of $10$; the k-th curve
from the top is rescaled with a factor $10^{-k}$.
The data are from \cite{ZEUS1,H11,NMC,H1}.
\item{{\bf Fig.\,5}} 
{\bf a.} Diffractive dissociaton of the photon
in photon-proton scattering due to the Pomeron exchange.\\
{\bf b.} Triple-regge diagram for hard diffractive scattering 
in the small $\beta$ region.
\item{{\bf Fig.\,6}} 
{\bf a.} Different parametrizations for the light quark
distribution of the Pomeron, $\Sigma_P(\beta)$,
at initial scales $Q_0^2$.
Solid line: this paper with $n_g=-0.5$.
Dotted line: this paper with $n_g=-0.9$.
Dot-dashed line: GS \cite{Gehrman}.
Dashed line: GK \cite{Golec}.\\
{\bf b.} Different gluonic distributions
in the Pomeron at initial scales $Q_0^2$.
Solid line: this paper with $n_g=-0.5$.
Dotted line: this paper with $n_g=-0.9$.
Dot-dashed line: GS \cite{Gehrman}.
Dashed line: GK \cite{Golec}.\\
{\bf c.} Predictions for the light quark singlet distributions       
at the factorization scale by different models.
Solid line: this paper with $n_g=-0.5$.
Dotted line: this paper with $n_g=-0.9$.
Dot-dashed line: GS \cite{Gehrman}.
Dashed line: GK \cite{Golec}.\\
{\bf d.} Predictions for gluon distributions 
at the factorization scale by  different models.
Solid line: this paper with $n_g=-0.5$.
Dotted line: this paper with $n_g=-0.9$.
Dot-dashed line: GS \cite{Gehrman}.
Dashed line: GK \cite{Golec}.
\item{{\bf Fig.\,7}} 
{\bf a.} Pomeron ``sea" charm contribution.\\
{\bf b.} Typical diagram that contributes to the 
``valence" charm component of the Pomeron.
\item{{\bf Fig.\,8}}
Pomeron structure function (upper curves)
and its charm contribution (lower curves)
predicted by different models as function of $\beta$
at different $Q^2$.
Solid line: this paper with $n_g=-0.5$.
Dot-dashed line: GS \cite{Gehrman}.
Dashed line: GK \cite{Golec}.\\
{\bf a.} $Q^2= 10$ GeV$^2$.\\
{\bf b.} $Q^2=100$ GeV$^2$.\\
{\bf c.} $Q^2=500$ GeV$^2$. 
\item{{\bf Fig.\,9}}
CKMT predictions with different $n_g$ for the Pomeron structure 
function (upper curves) and its charm contribution (lower curves).
Solid lines: $n_g=-0.5$. 
Dotted lines with $n_g=-0.9$. 
For every set the $Q^2$-values are from top to bottom:
$Q^2=500, 100, 10$ GeV$^2$.
\item{{\bf Fig.\,10}} 
Diffractive charm contribution to the total charm
production cross section for different values of $Q^2$;
from top to bottom $Q^2=0, 1.39, 2.47, 4.39, 7.81, 13.9, 24.7, 
43.9, 78.1$ GeV$^2$. 
Solid lines: $n_g=-0.5$.
Dotted lines: $n_g=-0.9$.
\item{{\bf Fig.\,11}} 
Cross sections for photo-and electroproduction of beauty
predicted by this paper for different values of $Q^2$;
from top to bottom $Q^2=0, 1.39, 2.47, 4.39, 7.81, 13.9, 24.7, 
43.9, 78.1$ GeV$^2$.
\item{{\bf Fig.\,12}} 
Beauty contribution to the Pomeron structure function
predicted by this paper at different values of $Q^2$:
from top to bottom $Q^2=500, 100, 10$ GeV$^2$.
Solid lines: $n_g=-0.5$.
Dotted lines: $n_g=-0.9$. 
\item{{\bf Fig.\,13}} 
Diffractive beauty contribution to the total beauty
production cross section predicted by this paper
for different values of $Q^2$;
from top to bottom $Q^2=0, 1.39, 2.47, 4.39, 7.81, 13.9, 24.7, 
43.9, 78.1$ GeV$^2$. 
Solid lines: $n_g=-0.5$.
Dotted lines: $n_g=-0.9$. 
\end{figures}
\newpage
\vspace*{4cm}
\centerline{\epsfysize=15cm \epsffile{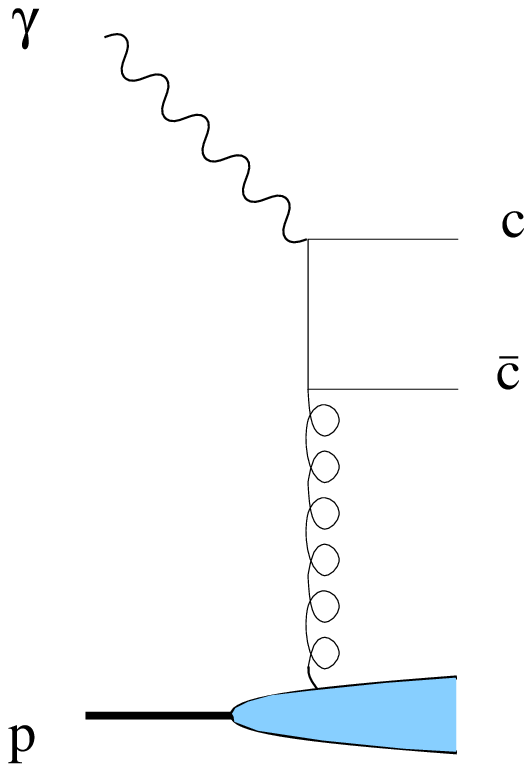}}
\nopagebreak
\begin{center}
{\LARGE {\bf Figure 1a} }
\end{center} \vspace*{0.5cm}
\newpage
\vspace*{4cm}
\centerline{\epsfysize=15cm \epsffile{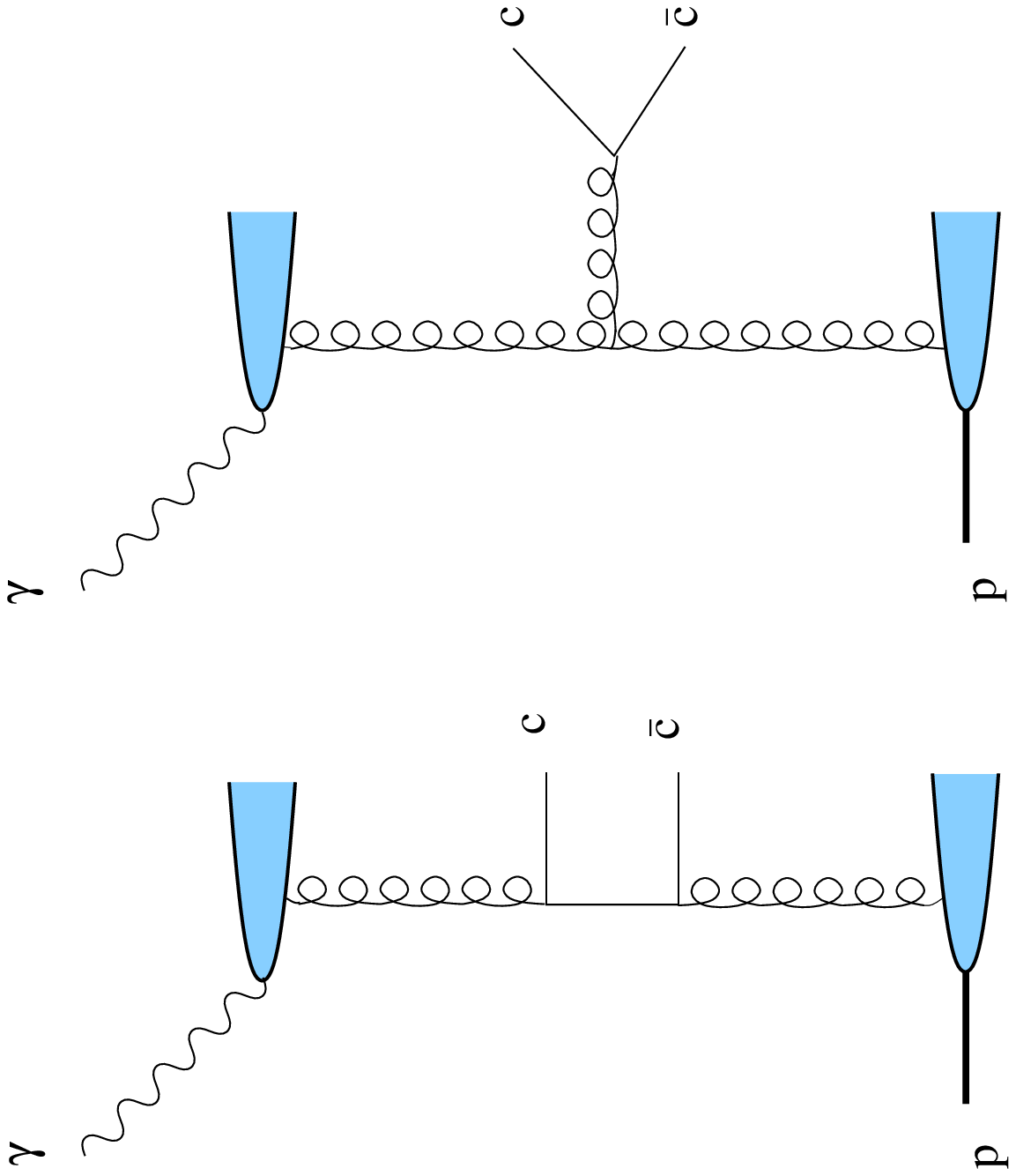}}
\nopagebreak
\begin{center}
{\LARGE {\bf Figure 1b} }
\end{center} \vspace*{0.5cm}
\newpage
\vspace*{4cm}
\centerline{\epsfysize=15cm \epsffile{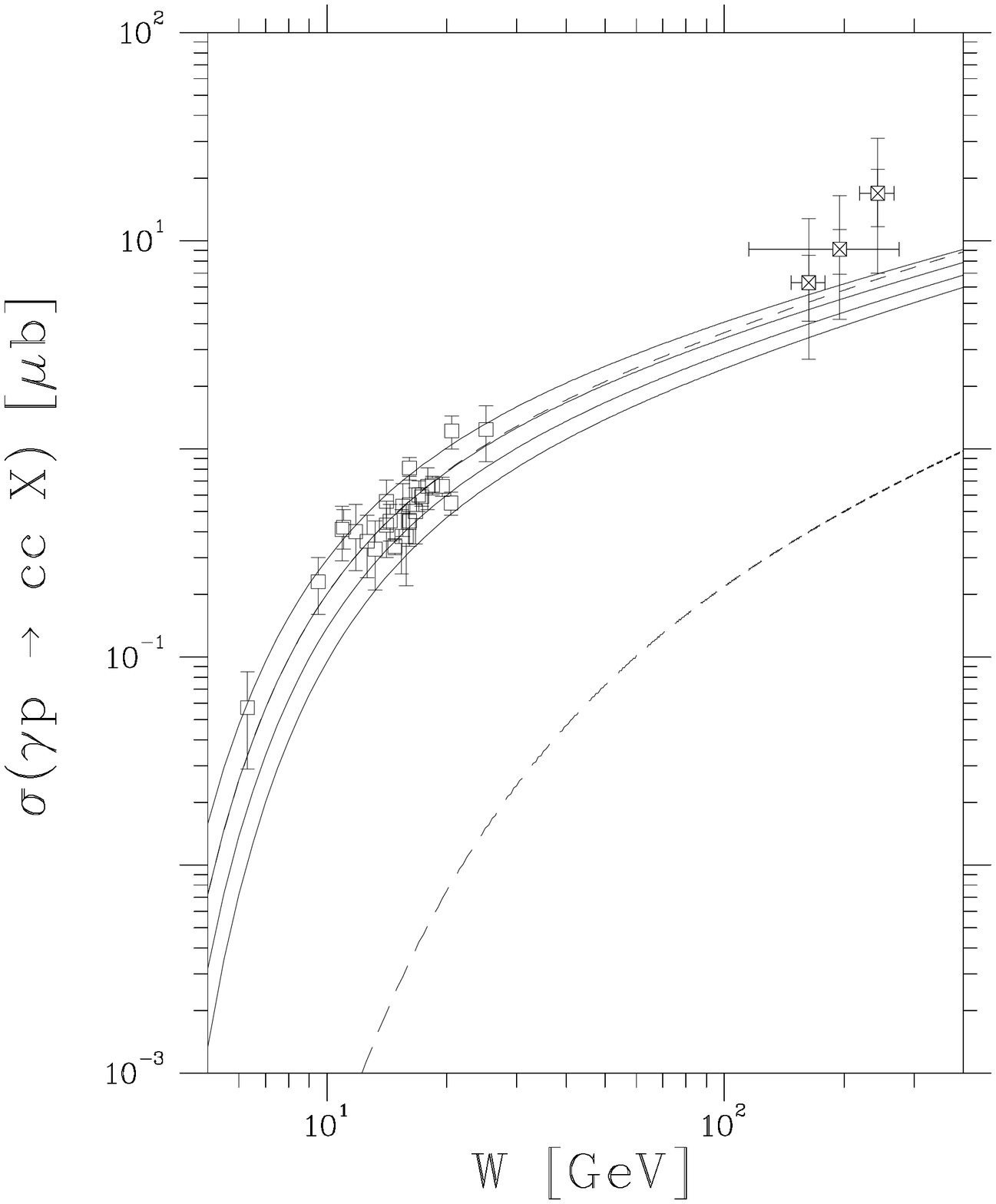}}
\nopagebreak
\begin{center}
{\LARGE {\bf Figure 2} }
\end{center} \vspace*{0.5cm}
\newpage
\vspace*{4cm}
\centerline{\epsfysize=15 cm \epsffile{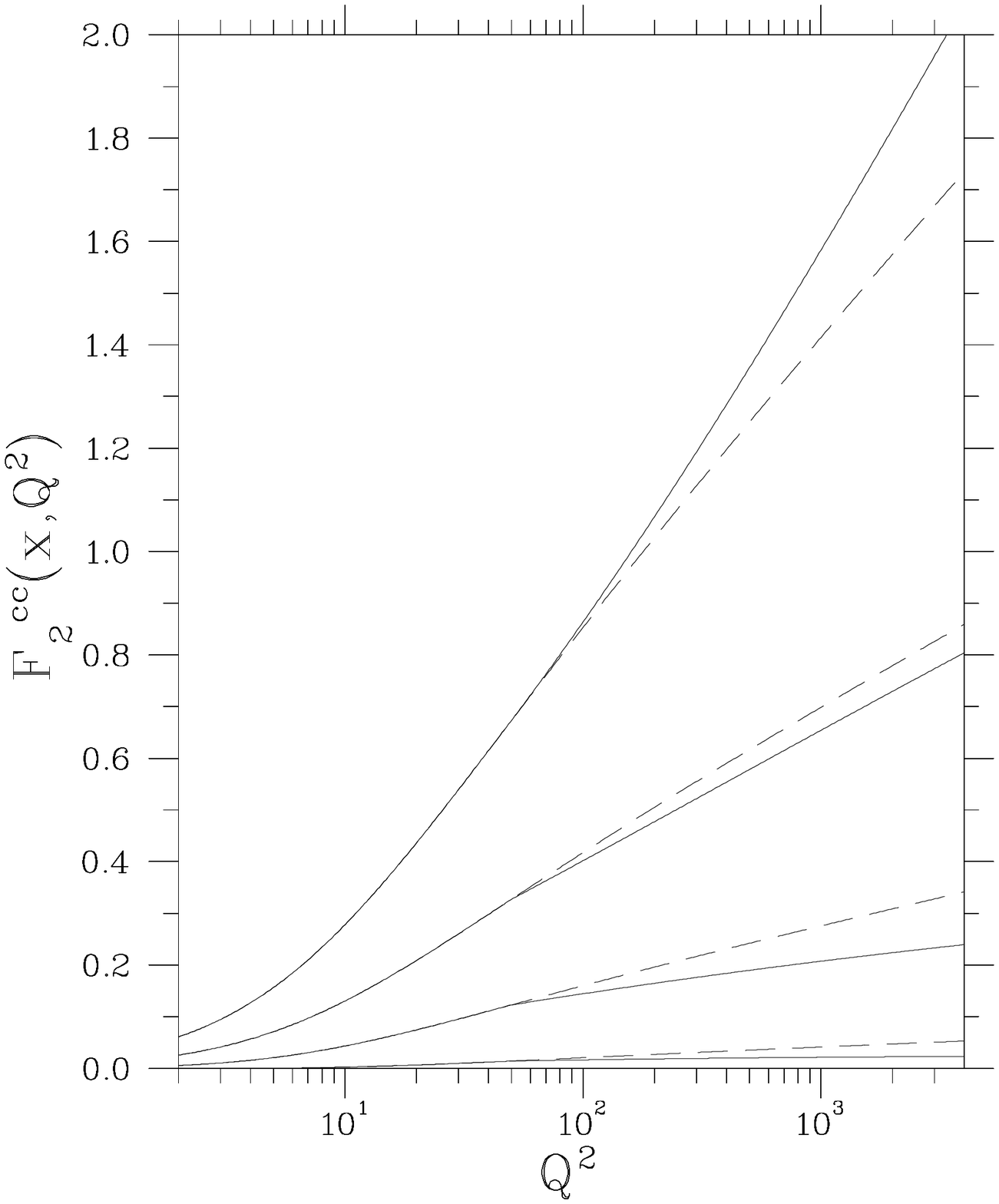}}
\nopagebreak
\begin{center}
{\LARGE {\bf Figure 3a} }
\end{center} \vspace*{0.5cm}
\newpage
\vspace*{4cm}
\centerline{\epsfysize=15cm \epsffile{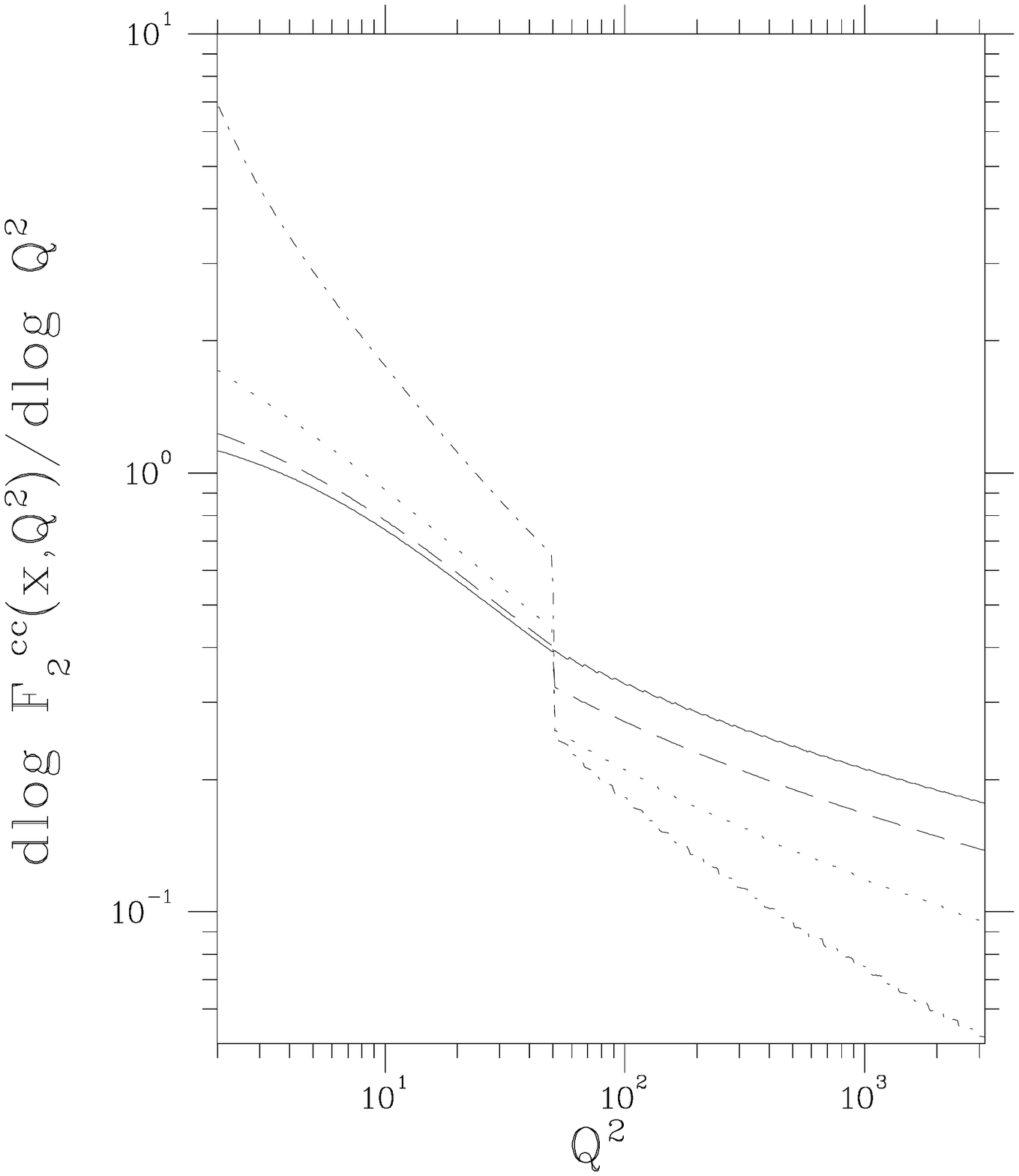}}
\nopagebreak
\begin{center}
{\LARGE {\bf Figure 3b} }
\end{center} \vspace*{0.5cm}
\newpage
\vspace*{4cm}
\centerline{\epsfysize=15cm \epsffile{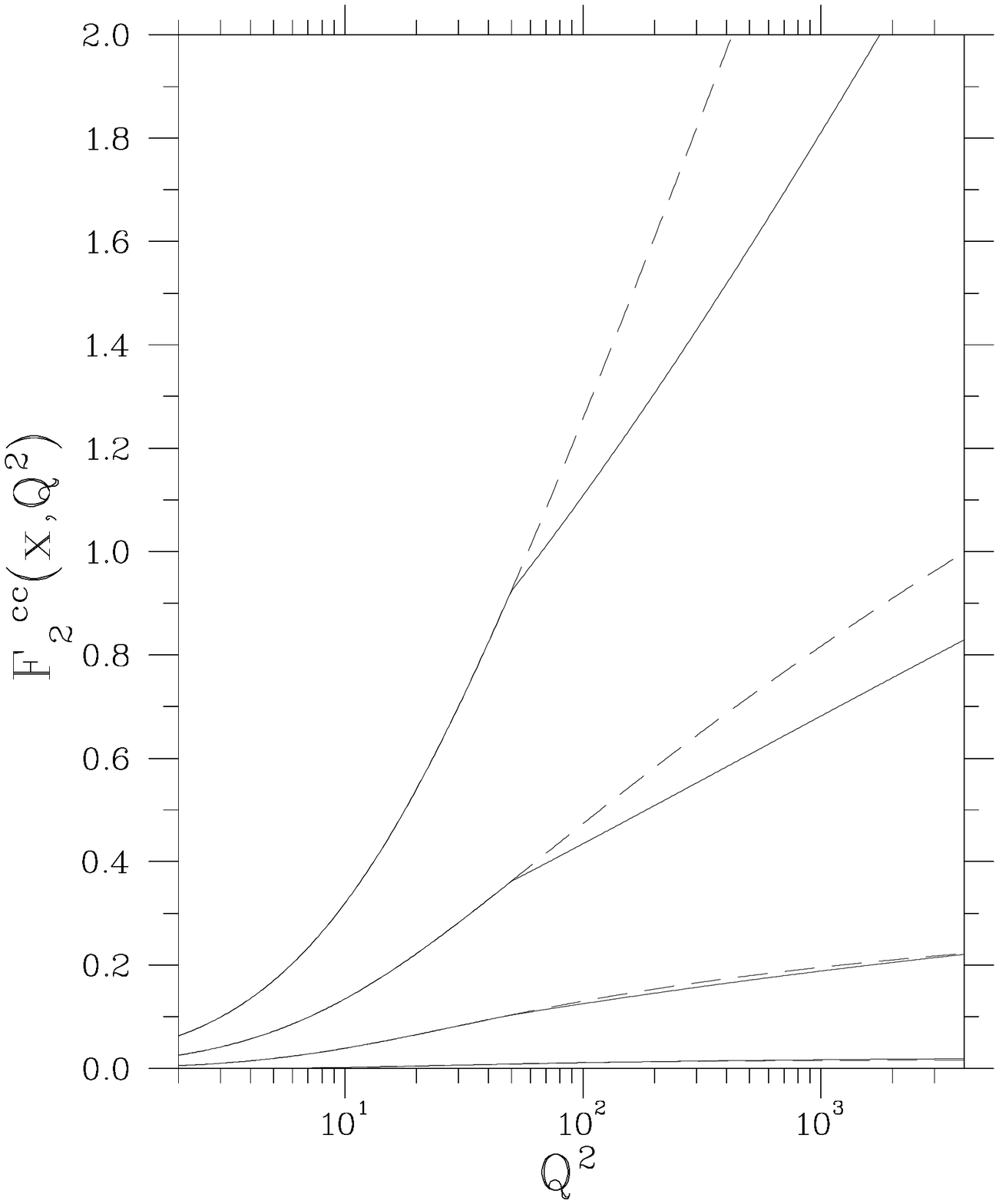}}
\nopagebreak
\begin{center}
{\LARGE {\bf Figure 3c} }
\end{center} \vspace*{0.5cm}
\newpage
\vspace*{4cm}
\centerline{\epsfysize=15 cm \epsffile{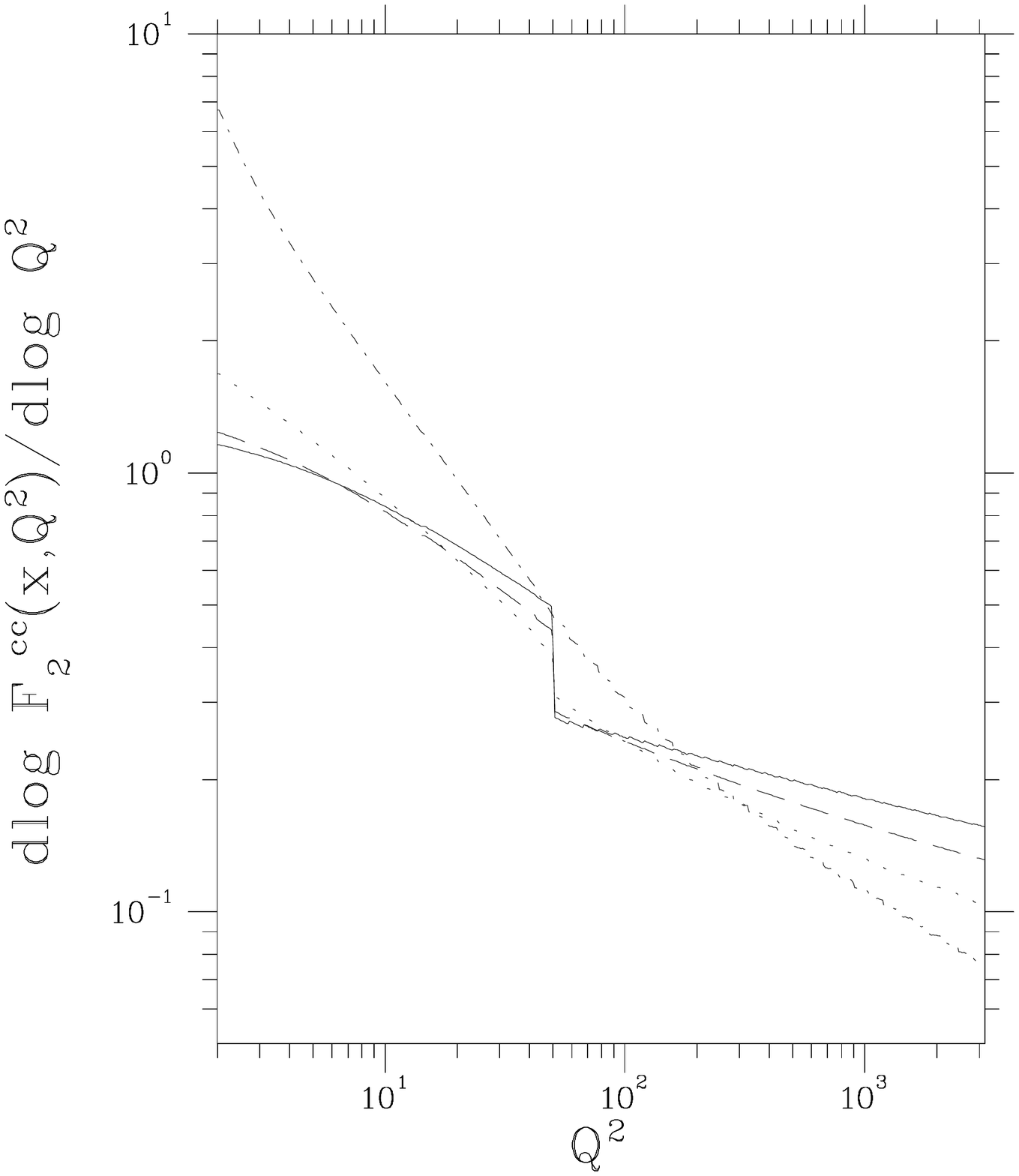}}
\nopagebreak
\begin{center}
{\LARGE {\bf Figure 3d} }
\end{center} \vspace*{0.5cm}
\newpage
\vspace*{4cm}
\centerline{\epsfysize=15cm \epsffile{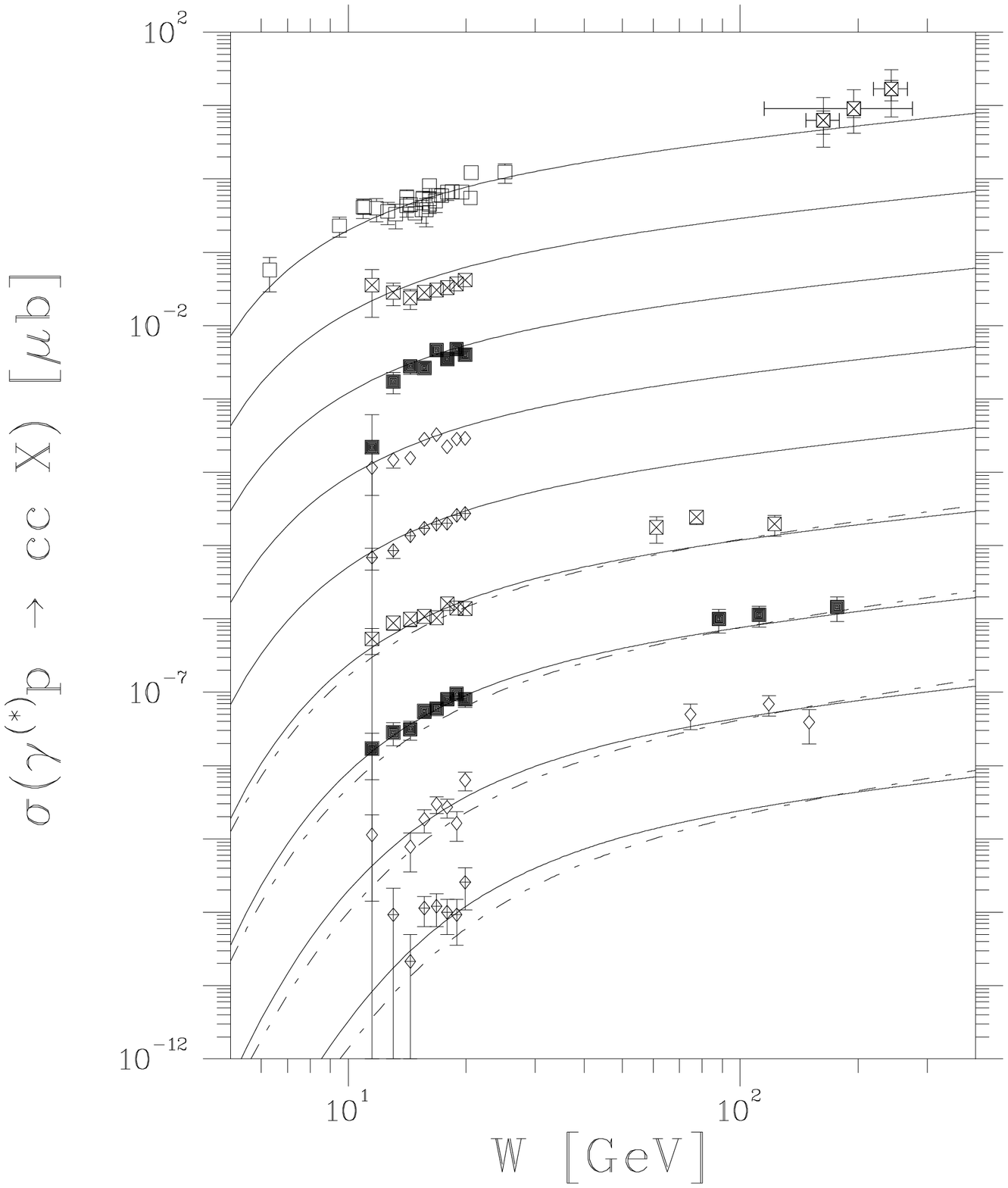}}
\nopagebreak
\begin{center}
{\LARGE {\bf Figure 4} }
\end{center} \vspace*{0.5cm}
\newpage
\vspace*{4cm}
\centerline{\epsfysize=15cm \epsffile{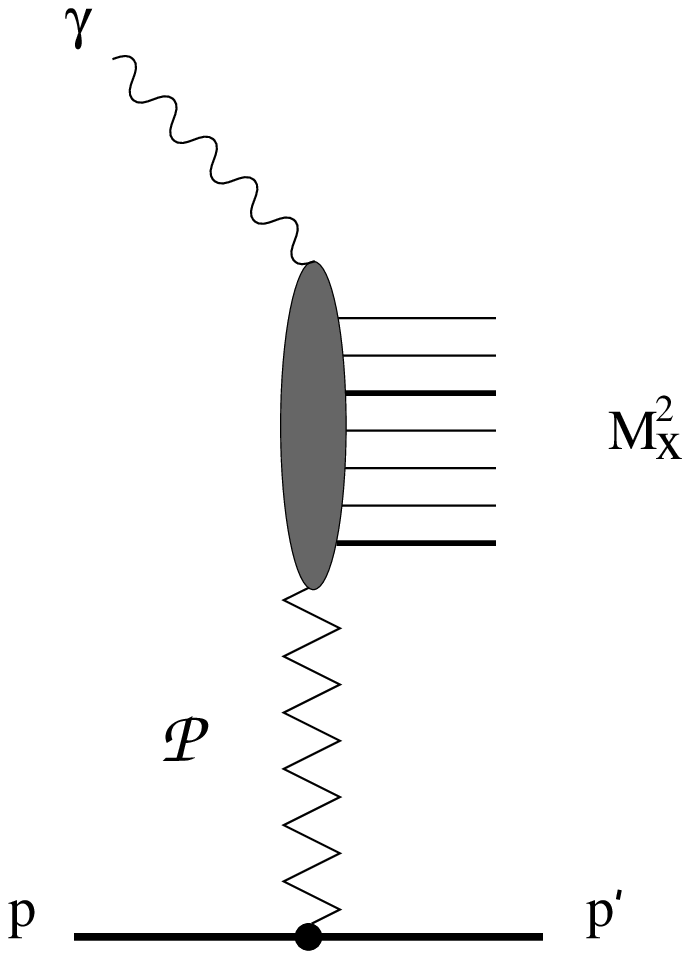}}
\nopagebreak
\begin{center}
{\LARGE {\bf Figure 5a} }
\end{center} \vspace*{0.5cm}
\newpage
\vspace*{4cm}
\centerline{\epsfysize=15 cm \epsffile{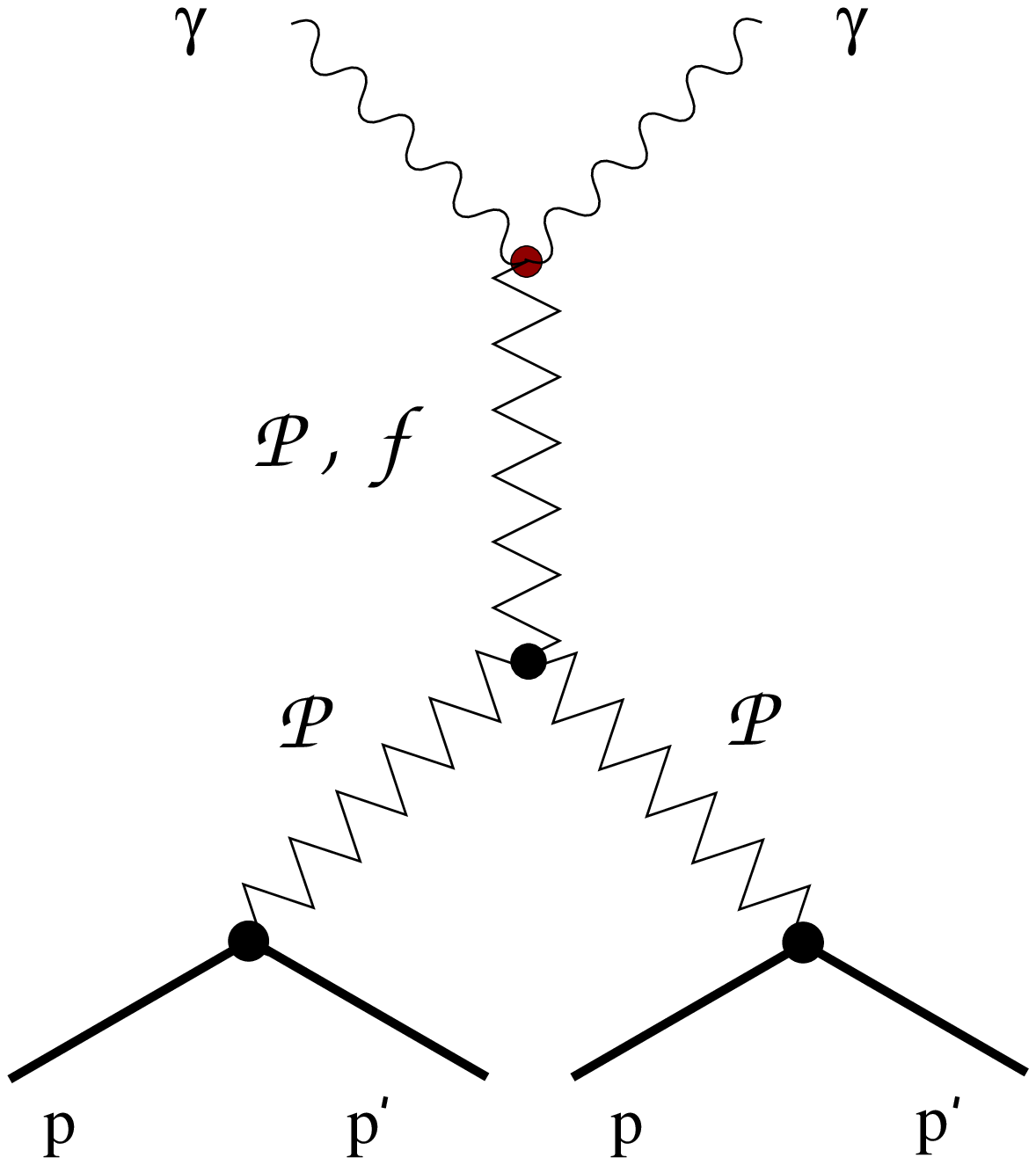}}
\nopagebreak
\begin{center}
{\LARGE {\bf Figure 5b} }
\end{center} \vspace*{0.5cm}
\newpage
\vspace*{4cm}
\centerline{\epsfysize=15cm \epsffile{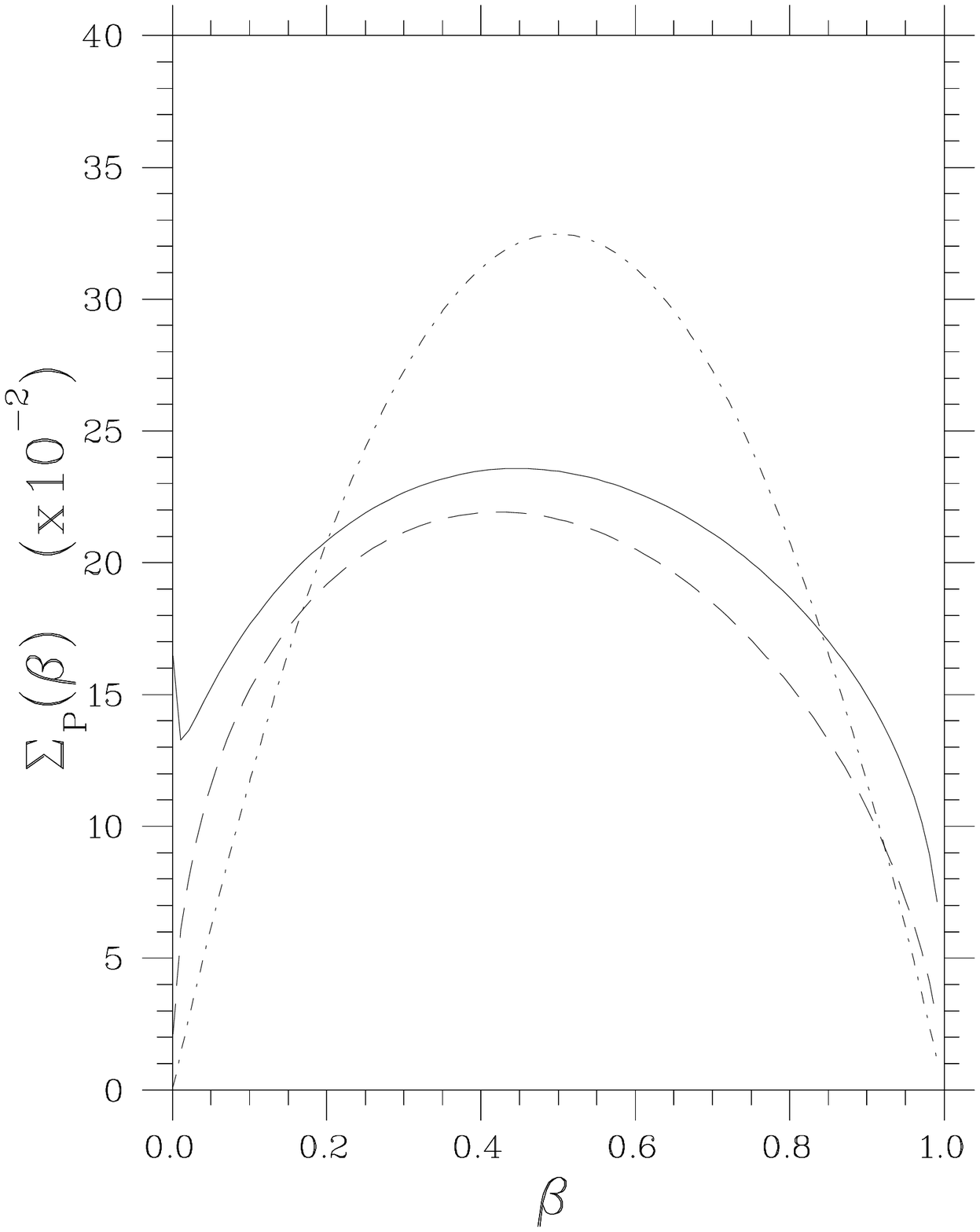}}
\nopagebreak
\begin{center}
{\LARGE {\bf Figure 6a} }
\end{center} \vspace*{0.5cm}
\newpage
\vspace*{4cm}
\centerline{\epsfysize=15cm \epsffile{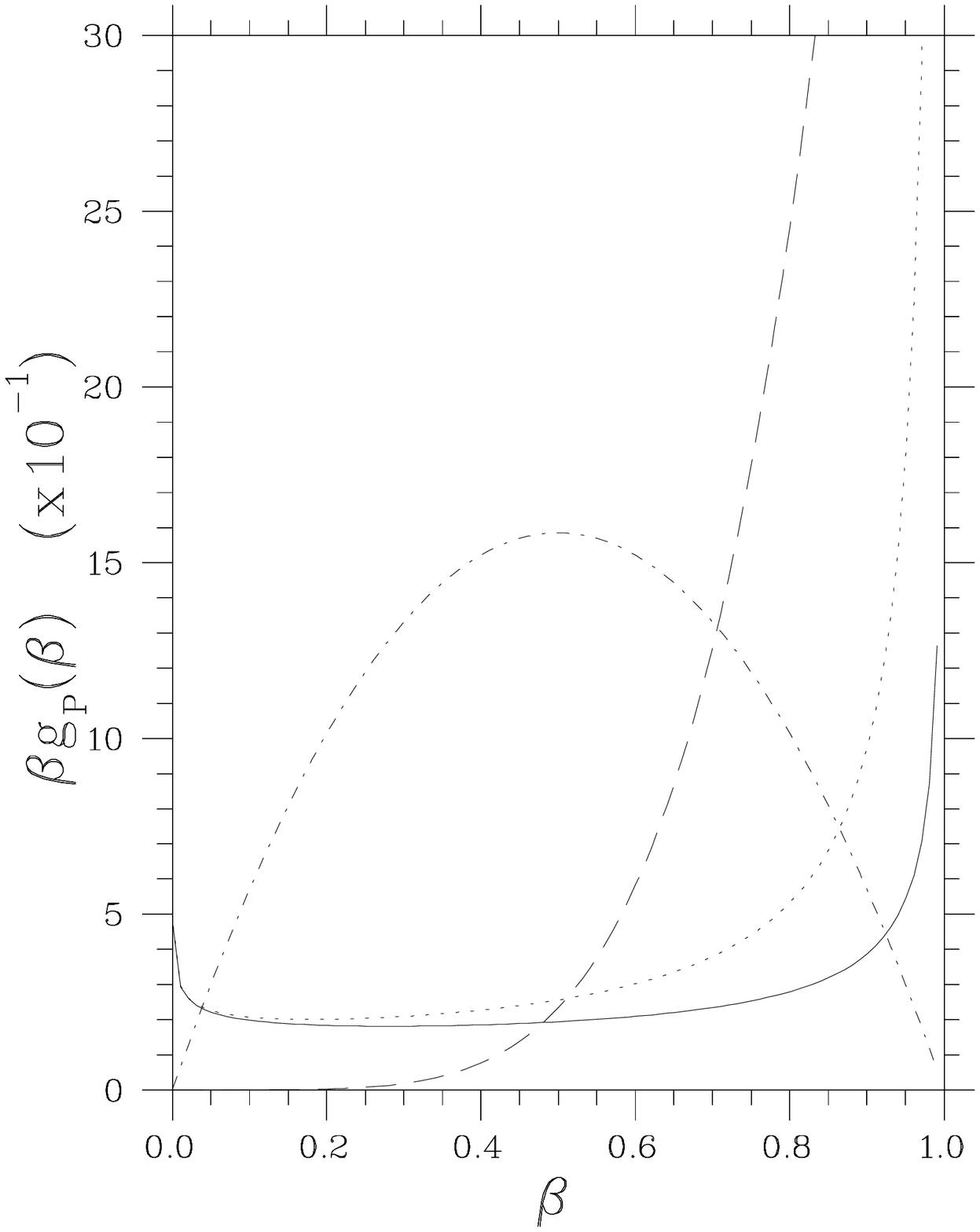}}
\nopagebreak
\begin{center}
{\LARGE {\bf Figure 6b} }
\end{center} \vspace*{0.5cm}
\newpage
\vspace*{4cm}
\centerline{\epsfysize=15 cm \epsffile{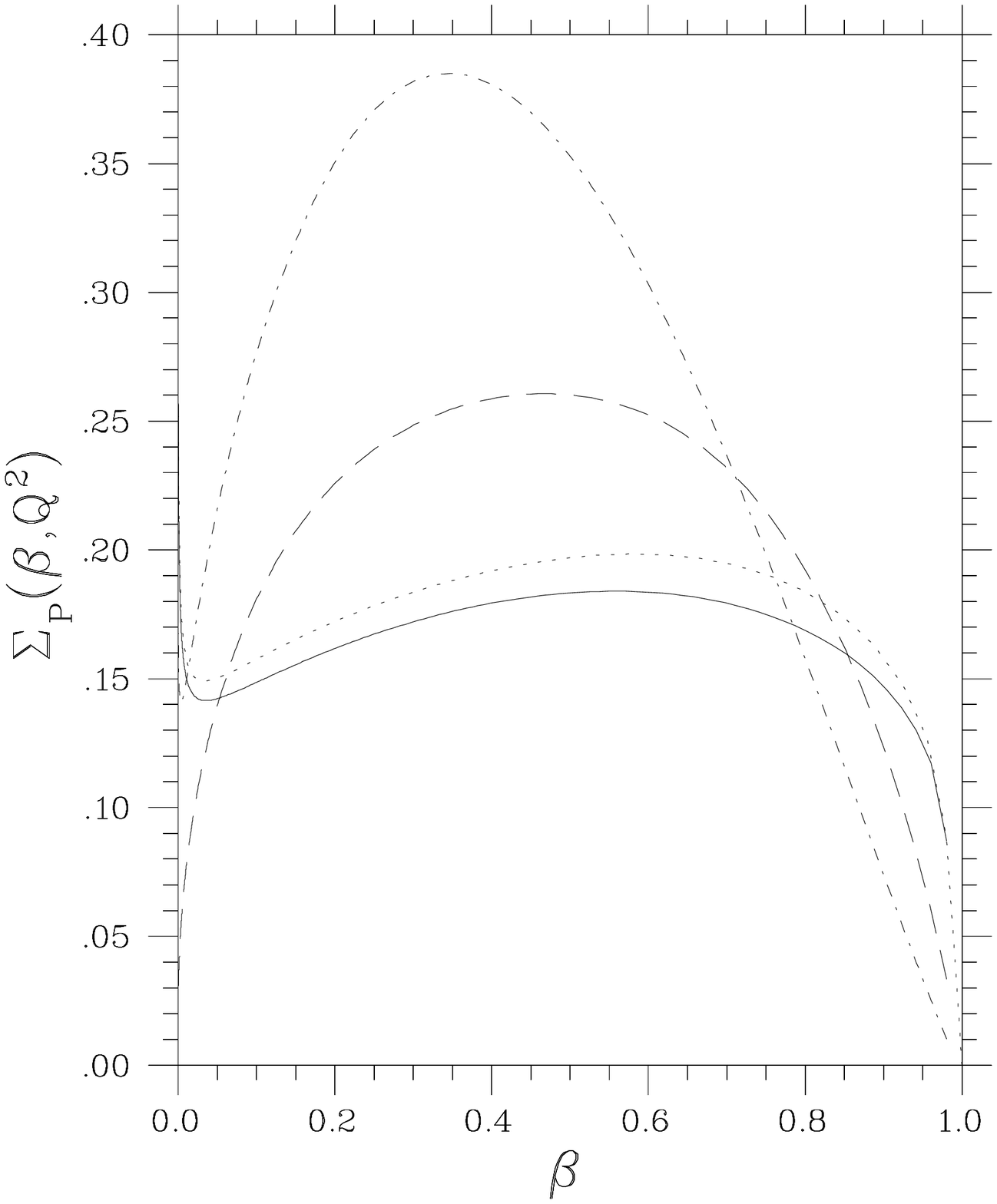}}
\nopagebreak
\begin{center}
{\LARGE {\bf Figure 6c} }
\end{center} \vspace*{0.5cm}
\newpage
\vspace*{4cm}
\centerline{\epsfysize=15cm \epsffile{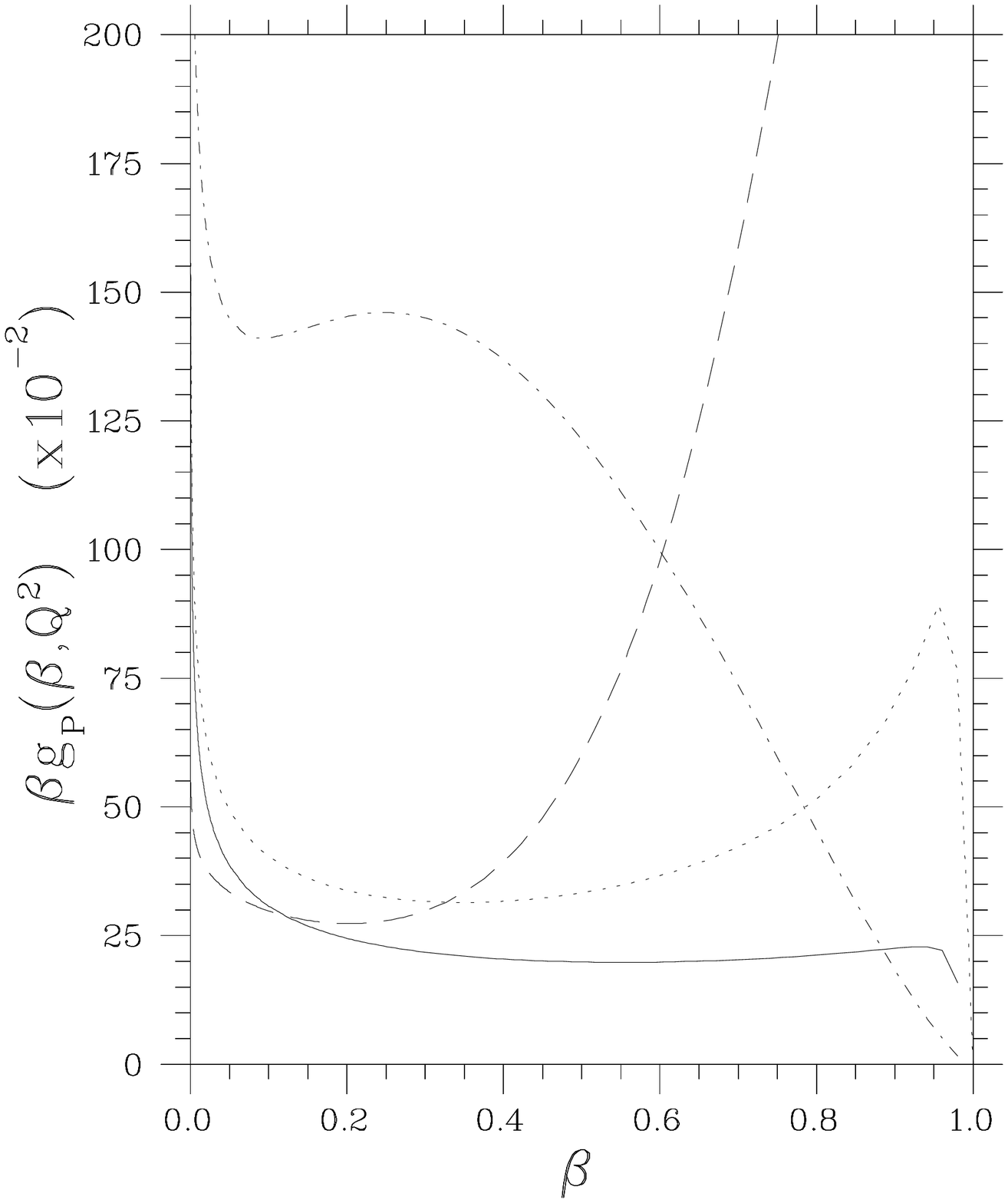}}
\nopagebreak
\begin{center}
{\LARGE {\bf Figure 6d} }
\end{center} \vspace*{0.5cm}
\newpage
\vspace*{4cm}
\centerline{\epsfysize=15 cm \epsffile{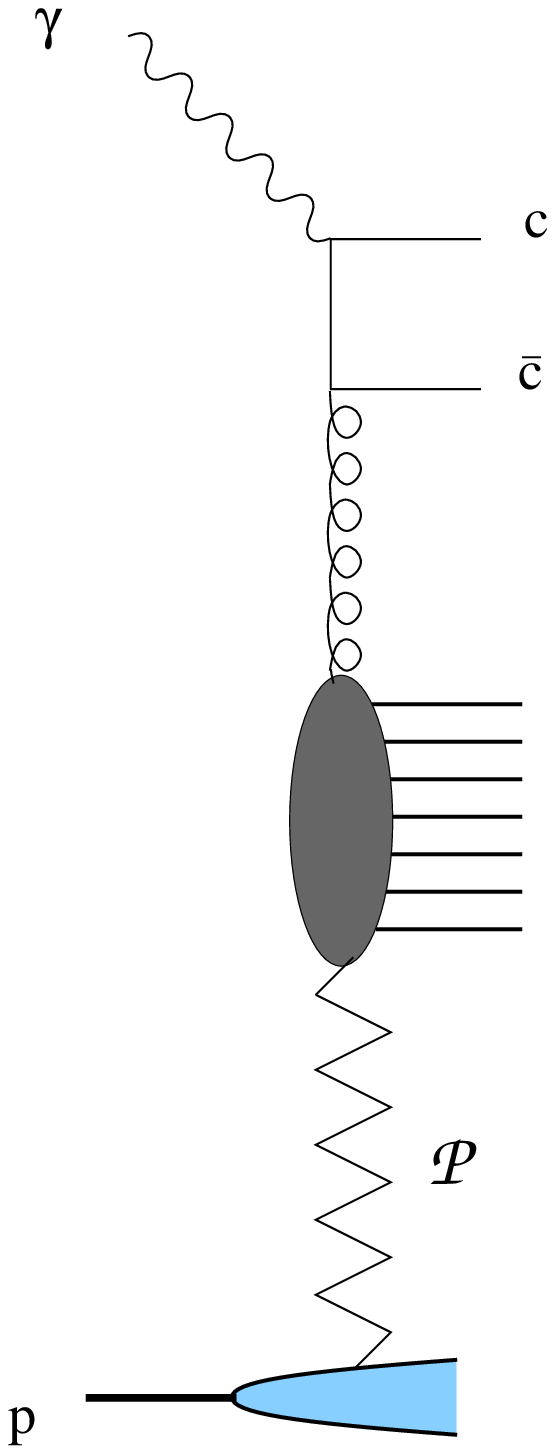}}
\nopagebreak
\begin{center}
{\LARGE {\bf Figure 7a} }
\end{center} \vspace*{0.5cm}
\newpage
\vspace*{4cm}
\centerline{\epsfysize=15cm \epsffile{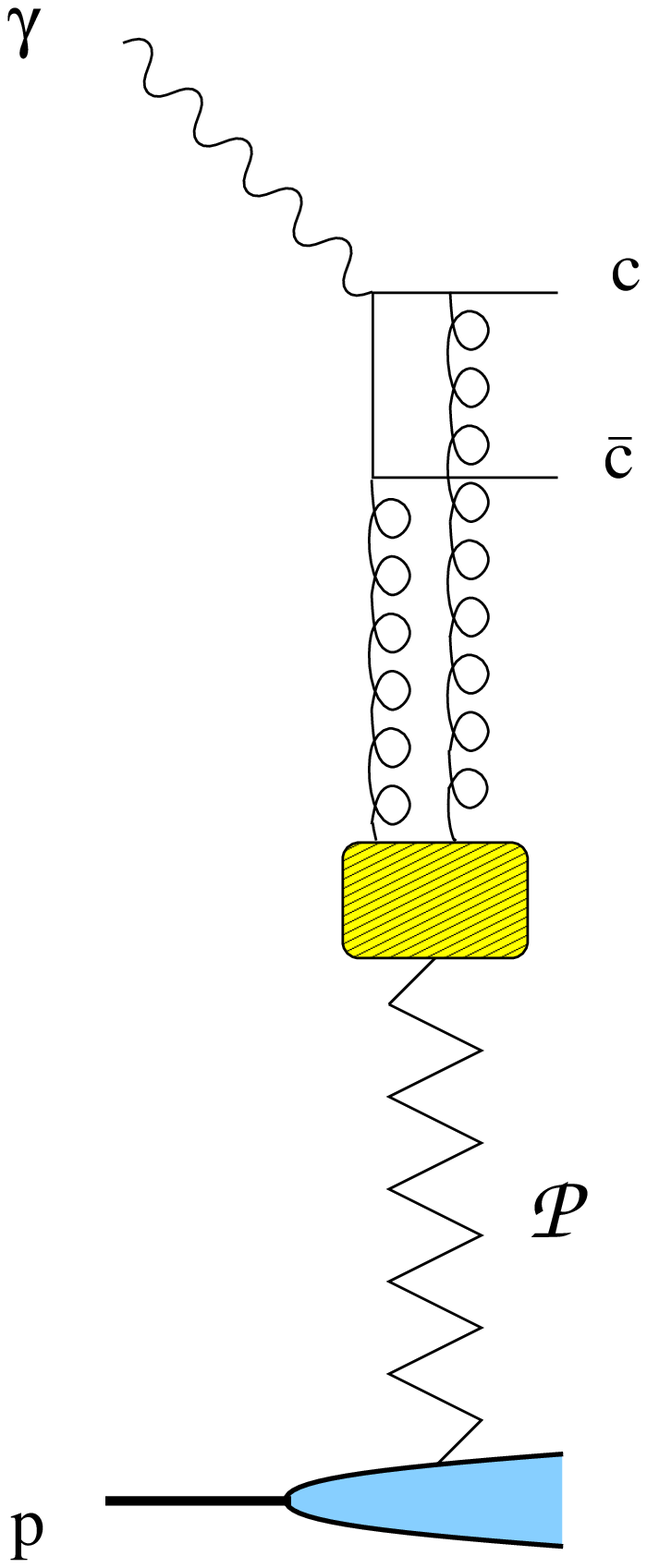}}\nopagebreak
\begin{center}
{\LARGE {\bf Figure 7b} }
\end{center} \vspace*{0.5cm}
\newpage
\vspace*{4cm}
\centerline{\epsfysize=15cm \epsffile{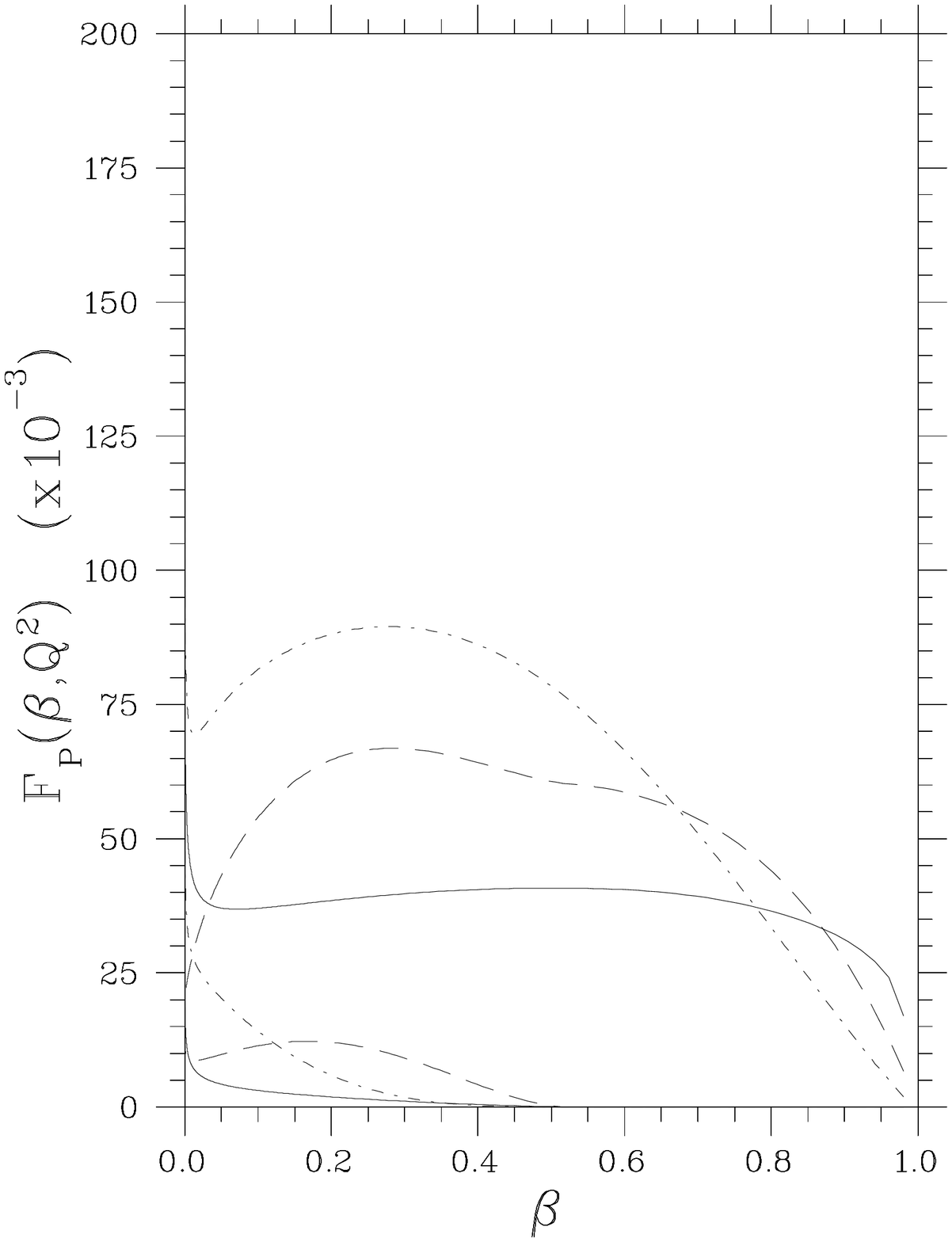}}
\nopagebreak
\begin{center}
{\LARGE {\bf Figure 8a} }
\end{center} \vspace*{0.5cm}
\newpage
\vspace*{4cm}
\centerline{\epsfysize=15cm \epsffile{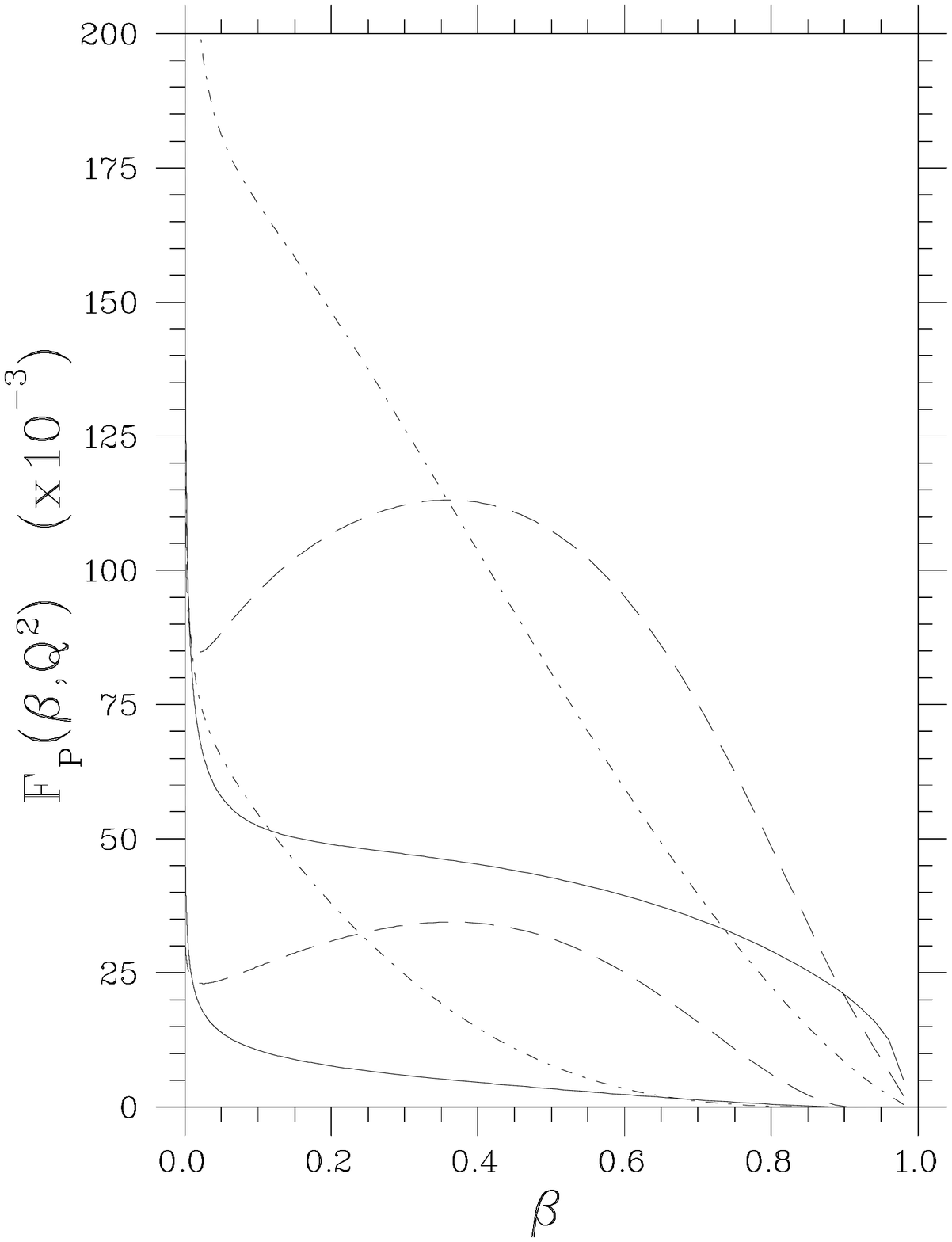}}
\nopagebreak
\begin{center}
{\LARGE {\bf Figure 8b} }
\end{center} \vspace*{0.5cm}
\newpage
\vspace*{4cm}
\centerline{\epsfysize=15cm \epsffile{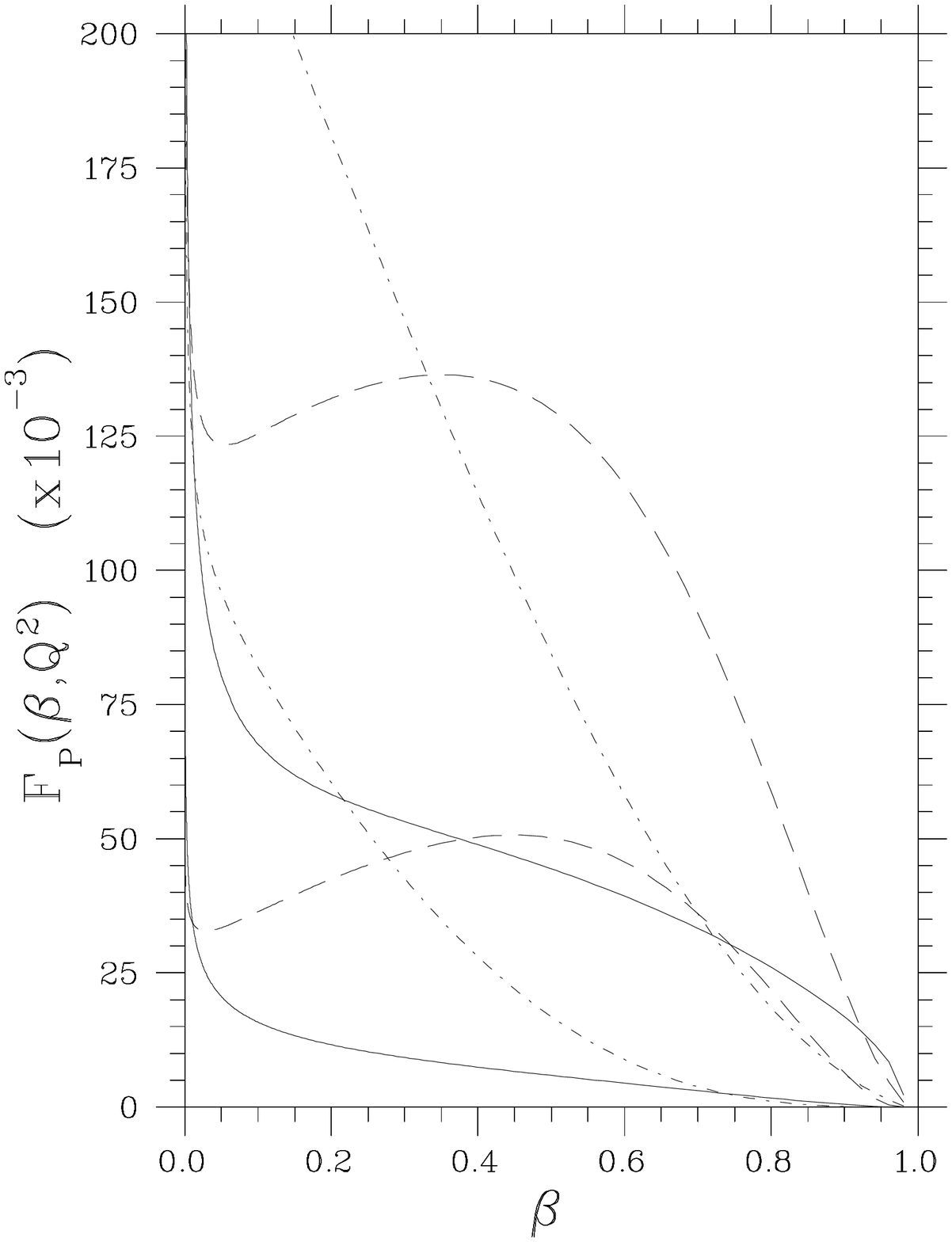}}
\nopagebreak
\begin{center}
{\LARGE {\bf Figure 8c} }
\end{center} \vspace*{0.5cm}
\newpage
\vspace*{4cm}
\centerline{\epsfysize=15 cm \epsffile{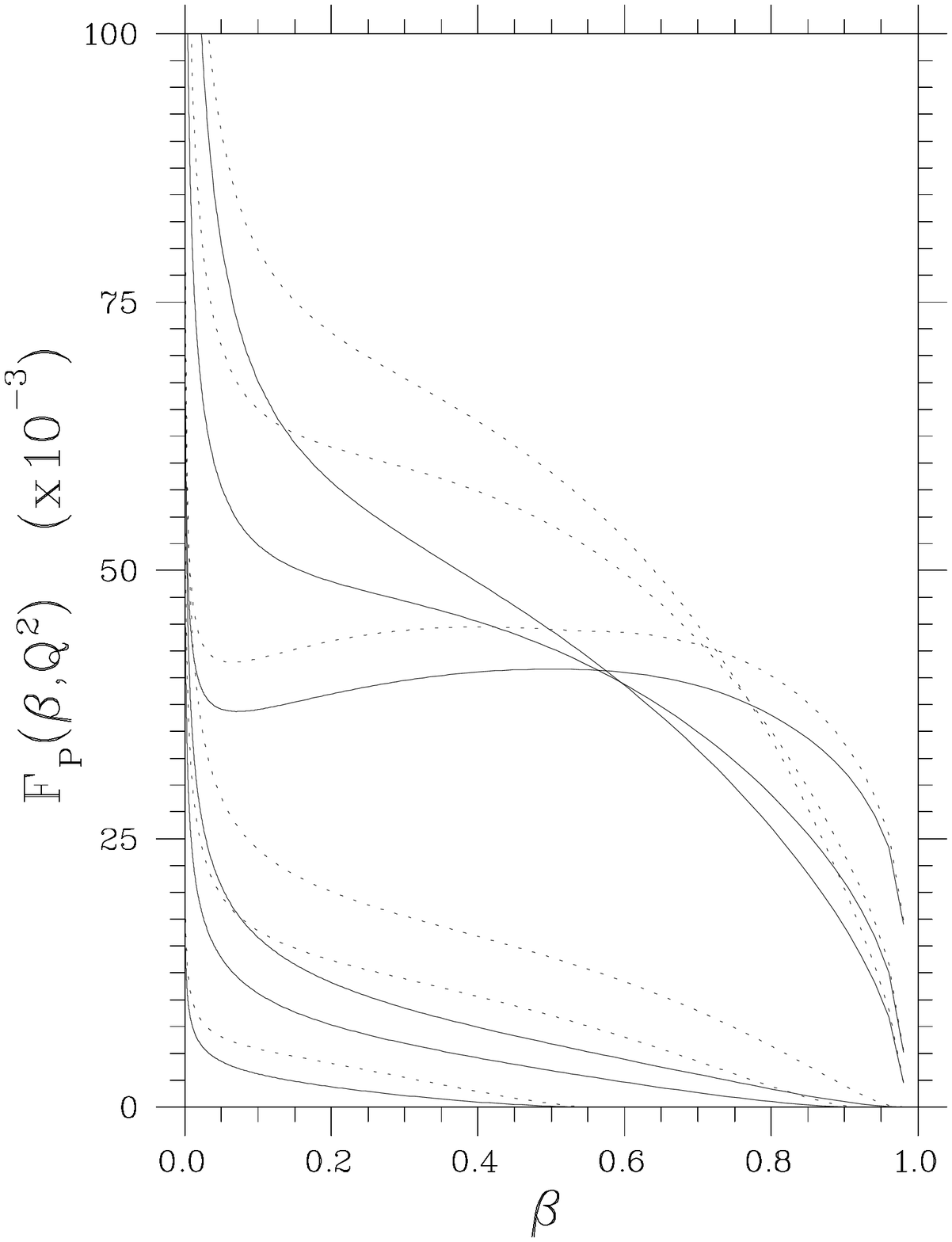}}
\nopagebreak
\begin{center}
{\LARGE {\bf Figure 9} }
\end{center} \vspace*{0.5cm}
\newpage
\vspace*{4cm}
\centerline{\epsfysize=15cm \epsffile{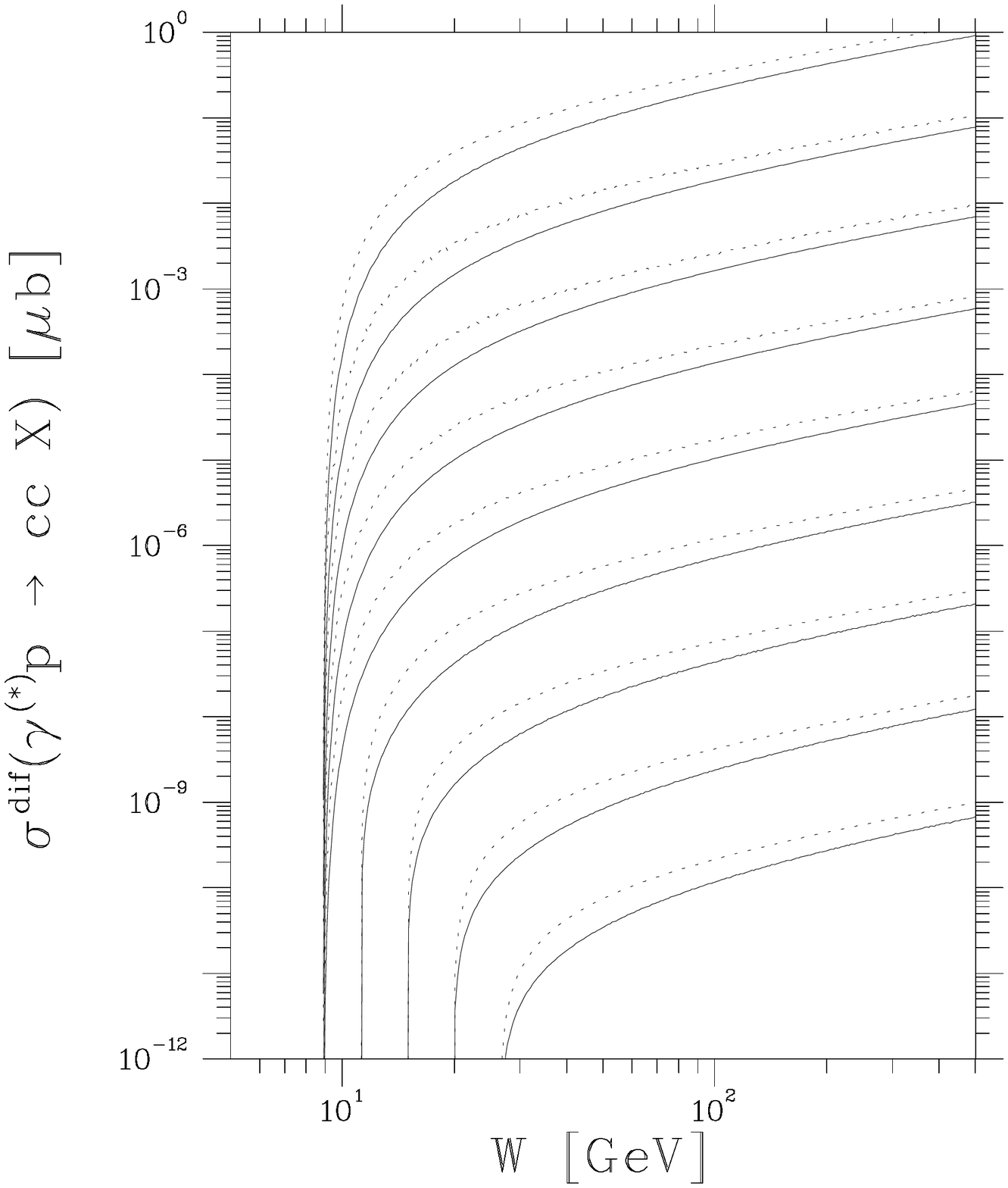}}
\nopagebreak
\begin{center}
{\LARGE {\bf Figure 10} }
\end{center} \vspace*{0.5cm}
\newpage
\vspace*{4cm}
\centerline{\epsfysize=15 cm \epsffile{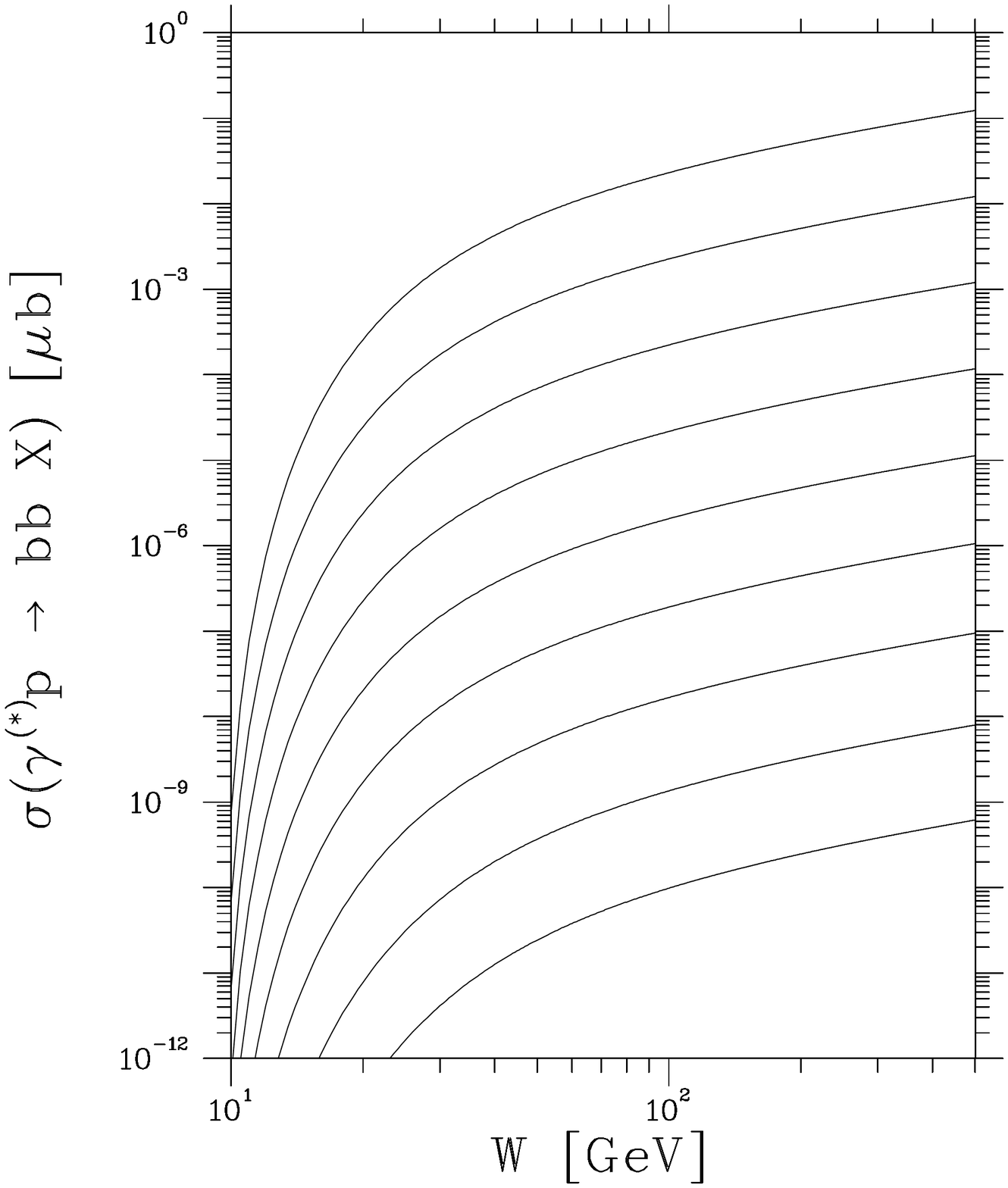}}
\nopagebreak
\begin{center}
{\LARGE {\bf Figure 11} }
\end{center} \vspace*{0.5cm}
\newpage
\vspace*{4cm}
\centerline{\epsfysize=15cm \epsffile{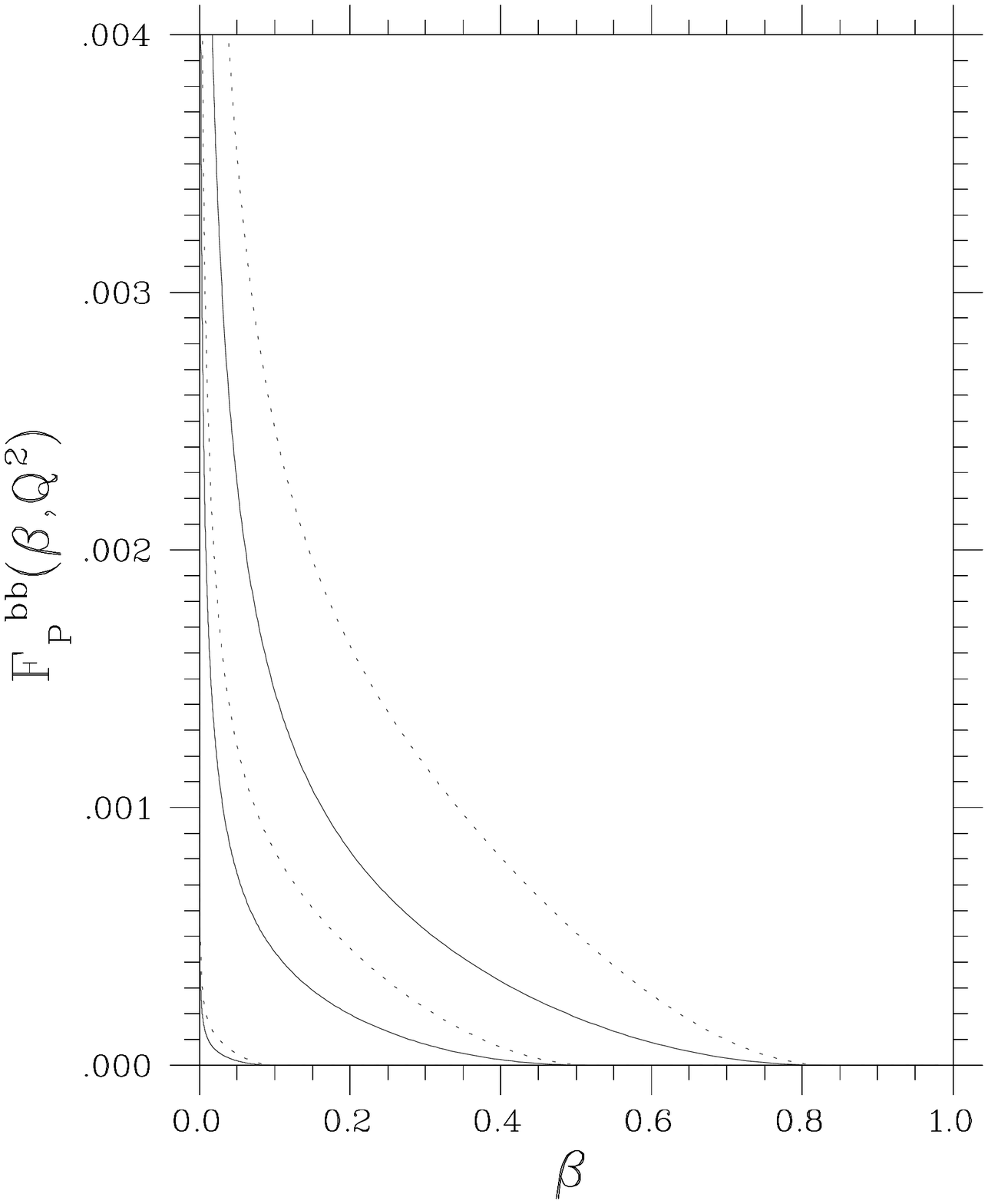}}
\nopagebreak
\begin{center}
{\LARGE {\bf Figure 12} }
\end{center} \vspace*{0.5cm}
\newpage
\vspace*{4cm}
\centerline{\epsfysize=15 cm \epsffile{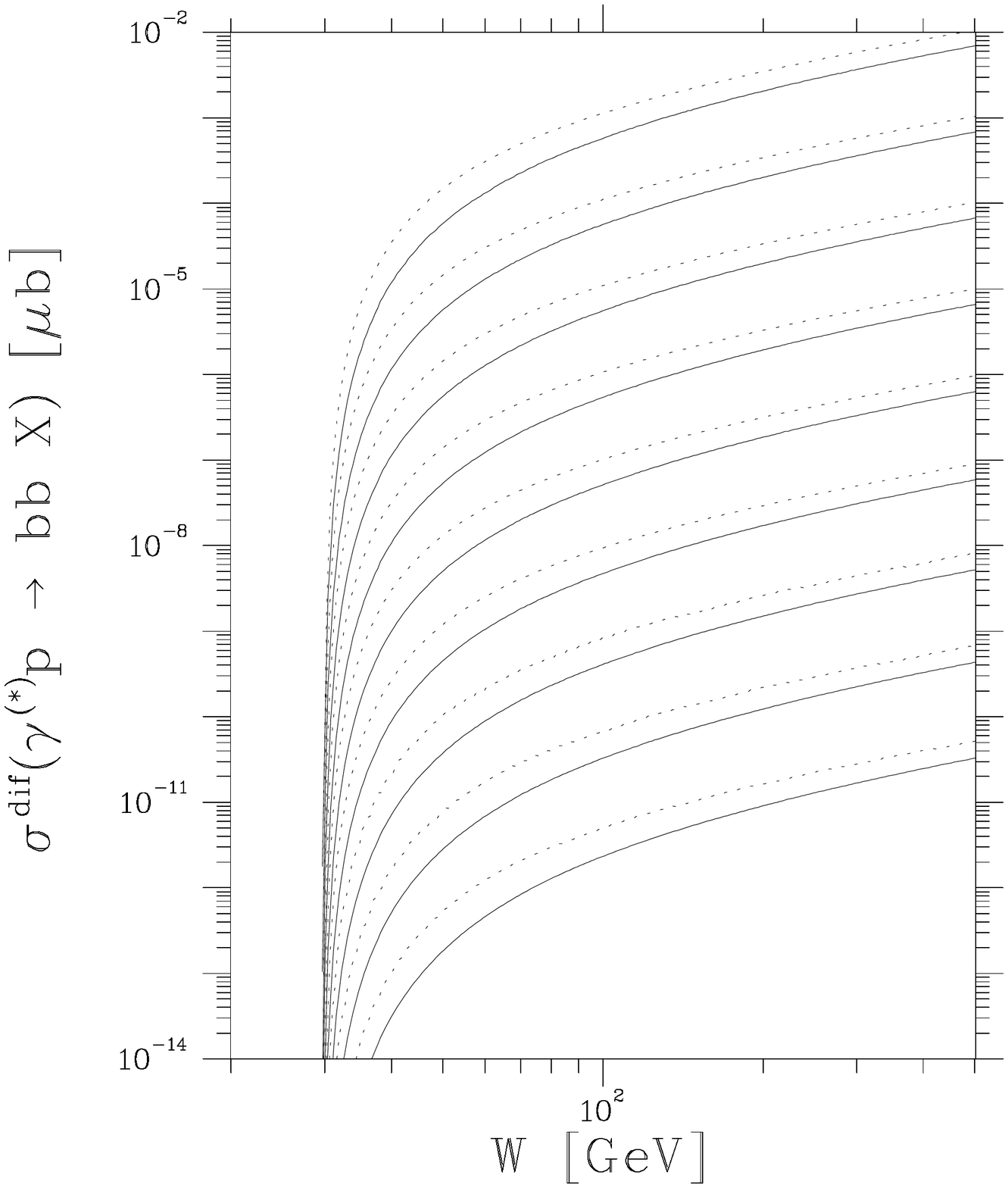}}
\nopagebreak
\begin{center}
{\LARGE {\bf Figure 13} }
\end{center} \vspace*{0.5cm}
\end{document}